\documentclass[preprint]{aastex}

\shortauthors{Sokal et al.}
\shorttitle{An Emerging WR Massive Star Cluster in NGC 4449}

\usepackage{url}
\usepackage{amsmath}
\usepackage{arydshln}

\begin{document}

\title{An Emerging Wolf-Rayet Massive Star Cluster in NGC 4449 }

\author{Kimberly R. Sokal\altaffilmark{1}}
\email{krs9tb@virginia.edu}
\author{Kelsey E. Johnson\altaffilmark{1,2}}
\author{R\'{e}my Indebetouw\altaffilmark{1,2}}
\author{Amy E. Reines\altaffilmark{2,3,4}}

\altaffiltext{1}{Department of Astronomy, University of Virginia, P.O. Box 3818, Charlottesville, VA 22903, USA}
\altaffiltext{2}{National Radio Astronomy Observatory, 520 Edgemont Road, Charlottesville, VA 22903, USA}
\altaffiltext{3}{currently Hubble Fellow at Department of Astronomy, University of Michigan, Ann Arbor, MI 48109, USA}
\altaffiltext{4}{Einstein Fellow}

%
\newcommand{\ltsimeq}{\raisebox{-0.6ex}{$\,\stackrel{\raisebox{-.2ex}%
{$\textstyle<$}}{\sim}\,$}}
%
\newcommand{\gtsimeq}{\raisebox{-0.6ex}{$\,\stackrel{\raisebox{-.2ex}%
{$\textstyle>$}}{\sim}\,$}}

\begin{abstract}
We present a panchromatic investigation of the partially-embedded, emerging massive cluster Source 26 ($=$ S26) in NGC 4449 with optical spectra obtained at Apache Point Observatory and archival {\em Hubble}, {\em Spitzer}, and {\em Herschel}\thanks{{\it Herschel} is an ESA space observatory with science instruments provided by European-led Principal Investigator consortia and with important participation from NASA.} {\em Space Telescope} images. First identified as a radio continuum source with a thermal component due to ionized material, the massive cluster S26 also exhibits optical Wolf-Rayet (WR) emission lines that reveal a large evolved massive star population.  We find that S26 is host to $\sim$240 massive stars, of which $\sim$18 are Wolf-Rayet stars; the relative populations are roughly consistent with other observed massive star forming clusters and galaxies. We construct SEDs  over two spatial scales ($\sim$100 pc and $\sim$300 pc) that clearly exhibit warm dust and polycyclic aromatic hydrocarbon (PAH) emission. The best fit dust and grain models reveal that both the intensity of the exciting radiation and PAH grain destruction increase toward the cluster center. Given that the timescale of evacuation is important for the future dynamical evolution of the cluster, it is important to determine whether O- and WR stars can evacuate the material gradually before supernova do so on a much faster timescale.  With a minimum age of $\approx$ 3 Myr, it is clear that S26 has not yet fully evacuated its natal material, which indicates that unevolved O-type stars alone do not provide sufficient feedback to remove the gas and dust.  We hypothesize that the feedback of WR stars in this cluster may be necessary for clearing the material from the gravitational potential of the cluster. We find S26 is similar to Emission Line Clusters observed in the Antennae Galaxies and may be considered a younger analog to 30 Doradus in the LMC.

\end{abstract}


\section{\label{sec-intro} Introduction}

The energetics of galaxies are largely driven by the evolution of massive stars; and when clustered, the impact of these stars can be catastrophic. Massive stars modify their environment through strong, fast winds and eventual supernova explosions. These processes inject energy and distribute heavy elements, making massive stars critical to galaxy evolution \citep{mae94}. Massive stars are found in the highest concentrations in massive and super star clusters (SSCs), which host hundreds to thousands of massive stars in a few parsecs and form in the most intense regions of star formation in the universe. However, the physical conditions that produce SSCs, or even slightly smaller massive star clusters, remain uncertain, although it is apparent the feedback from the constituent massive stars will drive the cluster evolution. In this work, we present a massive star cluster undergoing a major transition that may develop our understanding of the interplay of massive stars and super star cluster evolution.

A  picture of super star cluster evolution has developed in which SSCs form in thick, dense envelopes of natal material similar to scaled up versions of single massive stars \citep{john02}. SSC evolution starts with a molecular cloud proto-cluster that begins to form stars, yet as these stars are still embedded in the natal cocoon, the early evolution is effectively obscured from view at many wavelengths. However, the massive stars within the embedded cluster begin to ionize the surrounding material. A number of these analogs to Ultra-Compact HII (UCHII) regions have been identified in other galaxies; these vastly scaled up systems are detected as radio continuum sources \citep[e.g. ][]{kj99, turn00, jk03, john04b, tsai06,rei08, john09, tsai09, av11,kep14} with a flat or inverted spectral index, indicative of thermal free-free emission from dense young HII regions.  \citet{kj99} dubbed these sources as Ultra-Dense HII regions (UDHIIs) \citep[or similarly ``supernebulae'' by ][]{turn00}. 

The massive stars will continue to evolve and proceed to evacuate the surrounding material. One possible example of this emerging evolutionary stage are emission line clusters (ELC). Identified in the Antennae galaxies, ELCs are a type of HII region that are younger versions of SSCs and exhibit broadened Br$\gamma$ line emission suggestive of massive stars evacuating their surroundings via wind \citep{gg07}. Finally, the massive stars will be revealed at optical wavelengths and regulate or halt star formation \citep{ag13}. This results in the final early evolutionary stage of SSCs as bright and blue optical clusters that are well studied with HST \citep{ws95}.  This last stage can be exemplified by the well-known region 30 Doradus (30 Dor) in the LMC, which is the closest SSC analog. The entire massive/super star cluster evolutionary sequence can be summarized as: protocluster $\rightarrow$ UDHII $\rightarrow$ emerging cluster (ELC) $\rightarrow$ SSC \citep[e.g. ][]{w14}.

Although some of the steps in the evolution have been outlined, ultimately the physical process of an UDHII region becoming a cleared-out, optical SSC is not yet well understood. For instance, cluster age measurements are not feasible until the stars are optically visible after clearing embedding material and the most massive stars have started to evolve off the main sequence. Population comparisons suggest that the UDHII phase lasts for $<$ 1 Myr \citep{kj99}, however,  a radio and optical study of NGC 4449 revealed that some clusters may in fact remain embedded up to 5 Myr \citep{rei08}. As for the evacuation process, many models assume an instantaneous removal at some given age \citep[such as in ][]{pfa13}. This picture is too simplified, as the rate of removal will surely change the fate of the cluster \citep{bau07}. Even accounting for the rate of removal is not enough: as shown  by \citet{pfa13}, a scenario simply comparing SFE or expulsion timescales is too limited to describe a cluster's ability to stay bound and thus survive. 

Perhaps most paramount, the dominant mechanism responsible for the evacuation of the natal material is unclear.  Massive stars erode the obscuring envelope through a combination of stellar feedback processes including direct radiation from stars; pressure from cold, warm (the ionized HII region itself), and hot gas; dust processed IR radiation; protostellar winds and jets; and stellar winds and supernovae \citep[e.g.][]{lop13}. Yet, the relative importance of these mechanisms in removing the natal material is under debate -- especially how these processes are coupled to the molecular cloud material \citep[as discussed in][]{rp13}.

\begin{figure}[t]
\includegraphics*[width=18cm,angle=0]{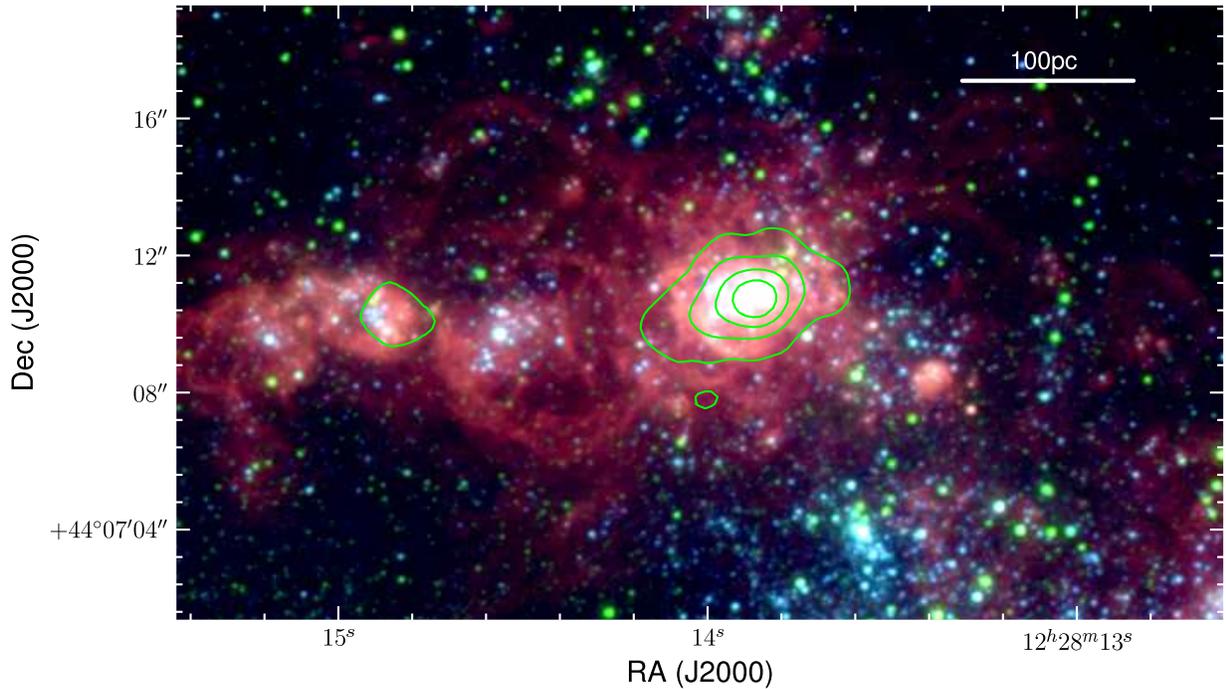}
\caption{\label{fig-hst} A {\em Hubble Space Telescope} rgb image (H$\alpha$, I, B) showing nebular emission surrounding a compact optical cluster in S26. The green contours show 3,4,5, and 6$\sigma$ emission at 3.6 cm \citep{rei08}; the 3$\sigma$ contour corresponds to a region with a radius of $\sim$50 pc. Color version available online.}
\end{figure}

\begin{figure}[t]
\includegraphics*[width=9cm,angle=0]{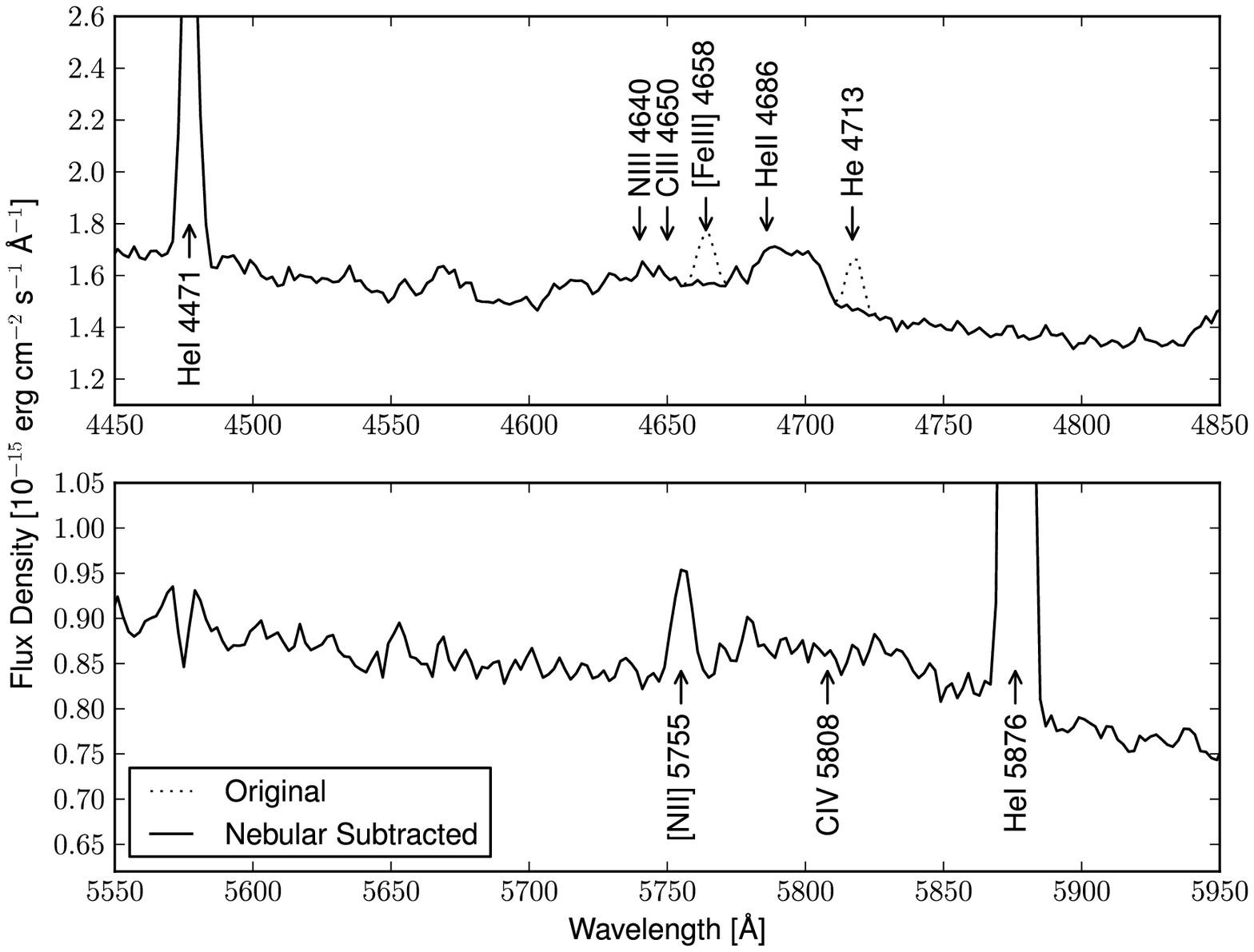}
\caption{\label{fig-wrspec} Flux calibrated spectra taken with DIS on the 3.5 m Telescope at Apache Point Observatory of the cluster S26 in NGC 4449, zoomed in on the broad Wolf-Rayet features. The dotted line shows the subtracted nebular features. Top- The ``blue bump'', a composite of broad lines at 4640 \AA, 4650 \AA, and 4686 \AA. Bottom- The broad ``red bump'' centered at 5808 \AA. Obvious nebular and WR features are labeled.}
\end{figure}

We have identified a young massive cluster in NGC 4449 that appears to be undergoing this major transition between evolutionary phases, and its stellar content is consistent with winds comprising a large component of the feedback driving the evolution.  The massive star cluster (roughly 50 pc; shown in Figure \ref{fig-hst}) catalogued by \citet{rei08} as Source 26 -- S26 hereafter-- in NGC 4449 simultaneously exhibits a thermal free-free radio emission component \citep{rei08} and optical spectra showing features that reveal the presence of Wolf-Rayet stars \citep[see Figure \ref{fig-wrspec} and][]{rei10} that begin to appear in $\sim$3 Myr \citep{con93}. Wolf-Rayet (WR) stars undergo rapid mass loss through fast winds with which they can drastically impact their environment. Thus, we hypothesize that we may have caught S26 in the act of breaking out of its natal cocoon, driven by the winds of the WR stars, providing an opportunity to observe the impact of evolved massive stars on their natal environment.

S26 is in a relatively low metallicity environment, and thus not only represents an important stage in cluster evolution, but also offers a window into a regime of star formation that is not well understood. The host galaxy NGC 4449 is an irregular Magellanic spiral that is close enough, at 3.9 Mpc \citep{ann08} where 1''$\sim$18 pc, to resolve individual star forming regions. Shown in Figure \ref{fig-irview}, the massive star cluster S26 can be found northward of the central part of the galaxy, at 12:28:13.86 $+$44:07:10.4 \citep{rei08}. The metallicity of the region has been measured from Z=0.004 \citep[$=$0.28 Z$_{\sun}$; ][]{rys11} to Z=0.0063 \citep[$=$0.44 Z$_{\sun}$; ][]{leq79}, which is comparable to the Large Magellanic Cloud (LMC) at Z=0.0068 \citep[O/H=8.37; ][]{rus90}. 

We present a multi-wavelength analysis of the massive star cluster S26 in NGC 4449. The optical spectra and archival infrared data and reduction are discussed in Section \ref{sec-obs}. We evaluate general properties of the region, such as extinction and age, in Section \ref{sec-opticalproperties}. We determine the massive star populations in Section \ref{sec-pops} and identify the thermal radio emission in Section \ref{sec-radiothermal}. In Section \ref{sec-dust}, we construct SEDs and find the best fit dust models.  In Section \ref{sec-discussion},  we put S26's massive star populations in context, discuss winds as possibly driving the evolution, and look at the similarities between S26 and 30 Doradus in the LMC. Finally, we present our conclusions briefly in Section \ref{sec-conclusions}.

\section{\label{sec-obs} Observations and Data Reduction}

\subsection{Optical Spectra from APO}

S26 in NGC 4449 was observed  on 2008 April 13 with the 3.5 m telescope at Apache Point Observatory \citep{rei10}, using the red and blue channels of the Dual Imaging Spectrograph (DIS) in low-resolution mode. The total exposure time was 30 minutes, producing a signal-to-noise ratio per pixel of $\sim$60 in the blue and $\sim$45 in the red continua. The spectrum has a resolution of $\sim$7 \AA\ over a wavelength range of $\sim$3800-9800 \AA\ and was reduced using IRAF and IDL routines. S26 was observed with a 1''.5 $\times$ 360'' slit and its spectrum was extracted from a 4''.4 window. The correction factor to account for slit loss is 1.9, determined by comparisons to HST photometry of a region of radius 3''.3 ($\approx$ 60 pc). Further details can be found in \citet{rei10}. 

The spectrum of the cluster S26 clearly displays typical Wolf-Rayet line features \citep[Fig. \ref{fig-wrspec}; also ][]{rei10}, explained in detail in Section \ref{sec-pops}. After the initial spectral reduction of \citet{rei10}, emission line features are further processed in this work using the IRAF SPLOT package. The blue bump WR feature, a composite feature of lines at 4650 and 4686 \AA, is analyzed after the prominent, superimposed nebular lines [Fe III] (4658 \AA) and HeI and [Ar IV] (4713 \AA) are subtracted (Fig. \ref{fig-wrspec}). The observed line flux, extinction corrected line fluxes (see Section \ref{sec-extinct}), and equivalent widths are given for emission lines in Table \ref{table-flux}. Results are similar to those in \citet{rei10} for the only presented line, H$\alpha$. 

\begin{deluxetable}{cclll}
\tabletypesize{\scriptsize}
\tablewidth{0pt} 
\tablecaption{\label{table-flux} Emission Line Properties of S26}
\tablehead{
	 \colhead{Wavelength}		&
           \colhead{Identification}             &
           \colhead{Observed Flux}                	&
           \colhead{Extinction Corrected Flux}                	&
           \colhead{EW}                   	\\  
           \colhead{(\AA)}		&
           \colhead{}                         	&
           \colhead{(10$^{-13}$ ergs cm$^{-2}$ s$^{-1}$)} 	&
           \colhead{(10$^{-13}$ ergs cm$^{-2}$ s$^{-1}$)} 	&
           \colhead{(\AA)}       
}
\startdata
   3835 &             H$\eta$ 	&     0.195 (0.013) &     0.306 (0.059) &     7.8 (0.7) \\ 
   3868 &              [NeIII] &     0.754 (0.033) &     1.18 (0.22) &    29.5 (2.7) \\ 
   3889 &            H$\zeta$ &     0.527 (0.023) &     0.82 (0.15) &    21.1 (1.5) \\ 
   3970 &         H$\epsilon$ &     0.607 (0.026) &     0.95 (0.17) &    25.1 (2.0) \\ 
   4076 &                [SII] &     0.019 (0.013) &     0.030 (0.021) &     0.8 (0.6) \\ 
   4102 &           H$\delta$ &     0.720 (0.023) &     1.12 (0.19) &    33.5 (1.5) \\ 
   4341 &           H$\gamma$ &     1.323 (0.042) &     2.03 (0.33) &    70.0 (4.1) \\ 
   4363 &               [OIII] &     0.066 (0.017) &     0.101 (0.031) &     3.6 (1.2) \\ 
   4471 &                  HeI &     0.116 (0.011) &     0.177 (0.033) &     7.1 (1.0) \\ 
   4650 &   [CIII] (blue bump) &     0.102 (0.018) &     0.155 (0.037) &     7.1 (1.3)\tablenotemark{a} \\ 
   4658 &              [FeIII] &     0.015 (0.001) &     0.0223 (0.0035) &     0.94 (0.02) \\ 
   4686 &     HeII (blue bump) &     0.077 (0.013) &     0.116 (0.026) &     5.3 (0.9)\tablenotemark{a}\\ 
   4711 &      [ArIV]/He I &     0.014 (0.001) &     0.0213 (0.0033) &     0.96 (0.02) \\ 
   4861 &            H$\beta$ &     3.00 (0.11) &     4.52 (0.68) &   185 (49) \\ 
   4959 &               [OIII] &     4.33 (0.24) &     6.50 (1.00) &   300 (200) \\ 
   5007 &               [OIII] &    13.40 (0.46) &    20.07 (2.96) &   940 (750) \\ 
   5755 &                [NII] &     0.009 (0.001) &     0.0125 (0.0017) &     1.0 (0.1) \\ 
   5808 &     [CIV] (red bump) &     0.040 (0.026) &     0.059 (0.039) &     4.9 (3.4)\tablenotemark{a}\\ 
   5876 &                  HeI &     0.348 (0.013) &     0.510 (0.067) &    43.6 (3.6) \\ 
   6300 &                 [OI] &     0.0410 (0.0035) &     0.0595 (0.0087) &     6.13 (0.72) \\ 
   6312 &               [SIII] &     0.0486 (0.0033) &     0.0705 (0.0096) &     7.25 (0.70) \\ 
   6548 &                [NII] &     0.101 (0.070) &     0.15 (0.10) &     8.7 (8.5) \\ 
   6563 &           H$\alpha$ &     9.60 (0.32) &    13.86 (1.66) &   580 (380) \\ 
   6584 &                [NII] &     0.350 (0.093) &     0.51 (0.15) &    46 (28) \\ 
   6678 &                  HeI &     0.106 (0.0051) &     0.153 (0.019) &    18.9 (1.6) \\ 
   6717 &                [SII] &     0.294 (0.024) &     0.423 (0.059) &    59 (27) \\ 
   6732 &                [SII] &     0.218 (0.023) &     0.314 (0.049) &    44 (23) \\ 
   7065 &                  HeI &     0.0778 (0.0066) &     0.112 (0.015) &    15.9 (2.7) \\ 
   7137 &              [ArIII] &     0.344 (0.013) &     0.493 (0.056) &    66.5 (8.6) \\ 
   7319 &                [OII] &     0.0728 (0.0039) &     0.104 (0.012) &    16.1 (1.5) \\ 
   7330 &                [OII] &     0.0584 (0.0034) &     0.083 (0.010) &    13.0 (1.3) \\ 
   7751 &              [ArIII] &     0.0827 (0.0048) &     0.117 (0.014) &    19.3 (2.3) \\ 
   9069 &               [SIII] &     0.974 (0.043) &     1.36 (0.14) &   500 (630) \\ 
   9532 &               [SIII] &     2.503 (0.086) &     3.49 (0.33) &    ...\tablenotemark{b} \\

\enddata

\tablecomments{
Tabulated quantities are: intrinsic wavelength, common identification, observed flux, extinction corrected flux, and equivalent width (EW). Uncertainties follow in parentheses. 
}
\tablenotetext{a}{The Wolf-Rayet features are broad.}
\tablenotetext{b}{Not well constrained.}

\end{deluxetable}

\subsection{Infrared Archival Data}

\begin{figure}[t]
\includegraphics*[width=12cm,angle=0]{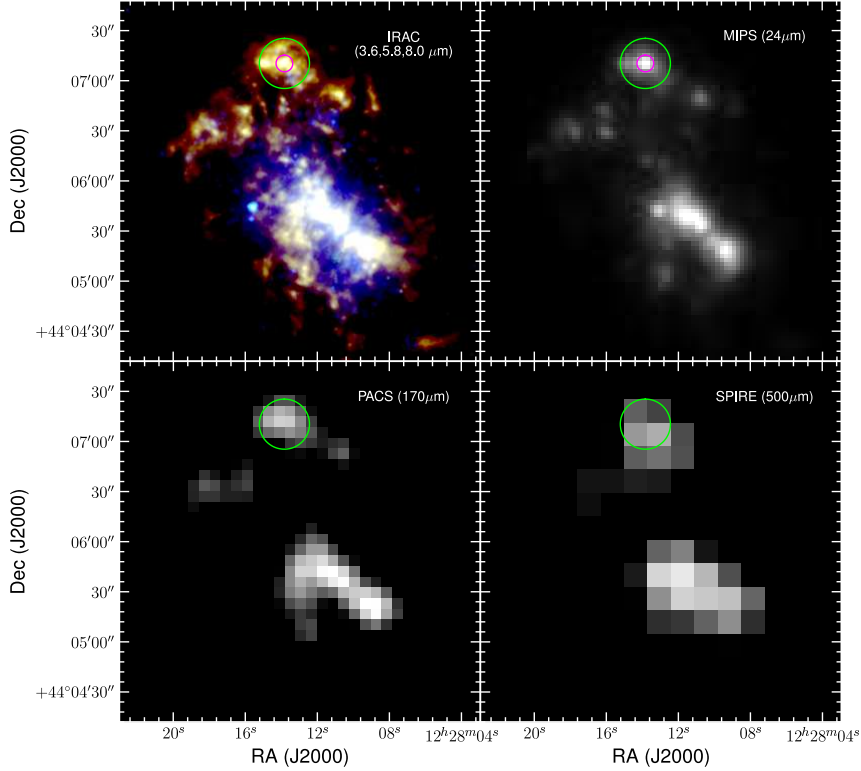}
\caption{\label{fig-irview} An infrared view of the massive cluster S26 in NGC 4449. From top left moving clockwise: an {\em Spitzer} IRAC rgb image (3.6 $\mu$m, 5.8 $\mu$m, and 8.0 $\mu$m),  {\em Spitzer} MIPS 24 $\mu$m,  {\em Herschel} PACS R (170 $\mu$m), and  {\em Herschel} SPIRE PLW (500 $\mu$m). The extraction regions for constructing the source SED (Fig. \ref{fig-sed}) are plotted in red (small 5''$\sim$100 pc circle) and green (large 15''$\sim$300 pc circle). S26 is a dominate source of emission even at the longest wavelengths. Color version available online. } \end{figure}

\subsubsection{{\em Spitzer} IRAC and MIPS}

Infrared data from the {\em Spitzer Space Telescope} \citep{wer04} were retrieved from the Spitzer Science Center Archive, consisting of InfraRed Array Camera (IRAC) \citep{faz04} and  Multiband Imaging Photometer for SIRTF (MIPS) \citep{riek04} imaging. NGC 4449 was imaged with IRAC (PI: G. Fazio) at 3.6 $\mu$m, 4.5 $\mu$m, 5.8 $\mu$m, and 8.0 $\mu$m \citep[FWHM of $\sim$1.9''; ][]{faz04} and MIPS (PI: R. Kennicutt) at 24 $\mu$m \citep[FWHM of $\sim$6''; ][]{riek04}. We analyze the post-basic calibrated data from IRAC and MIPS images, which were reduced with the Spitzer Science Center pipeline.  The flux calibration uncertainty is 2\% \citep{rea05} for IRAC and 4\% for MIPS \citep{eng07}.

\subsubsection{{\em Herschel} PACS and SPIRE }

The ESA Herschel Space Observatory \citep{pil10} has observed NGC 4449 as part of the Dwarf Galaxy Survey \citep{mad13}. NGC 4449 was observed with the Photodetector Array Camera and Spectrometer (PACS) \citep{pog10} at 70, 100, and 160 $\mu$m (FWHM of the PSF is 5.2, 7.7,  and 12.0'', respectively) and the Spectral and Photometric Imaging REceiver (SPIRE) \citep{grif10}at 250, 350, and 500 $\mu$m (FWHM of 18.2, 24.9, and 36.3'', respectively). 

We analyze the newly released `Level 2.5' data from PACS, using MadMap images to preserve extended structure \citep{rr13}. The PACS calibration uncertainty is of the order of 10\%, e.g. \citet{fri12}. For the SPIRE data, we analyze the `Level 2.0' and adopt a conservative uncertainty of 15\% on flux calibration, as according to the SPIRE observer's manual (\url{http://herschel.esac.esa.int/Docs/SPIRE/html/}) the overall calibration uncertainty for the SPIRE photometer is 7-15\%.  For both PACS and SPIRE datasets, we apply astrometric corrections to the data using the peak emission of the nucleus of NGC 4449.

\section{\label{sec-opticalproperties} General Properties of S26}

\subsection{\label{sec-extinct} Extinction}

The extinction of a star forming region is crucial for accurate line measurements, as well as information on the extent to which a source is embedded. Using nebular lines only measures extinction towards gas that is not very heavily extincted and thus can be biased low. We derive the extinction of S26 through optical nebular Balmer line and radio fluxes from \citet{rei08}. Extinction curves for the 30 Doradus region of the LMC \citep{mis99,fitz85} are used to convert the Balmer decrement to A$_\text{V}$, appropriate as the dust processing and the metallicity of the two regions are similar, roughly z=0.008 \citep{rus90}. Use of a Milky Way extinction curve changes the measured extinction by only $\sim$10\%. 

The total extinction of S26 is measured to be A$_\text{V}$ $=$ 0.41 and is used in correcting the optical line fluxes in Table 1. Subtraction of the Galactic foreground extinction results in an internal extinction of A$_\text{V, i}$ $=$ 0.35: along the S26 line of sight, the Milky Way galactic extinction is measured as E(B$-$V)$=$ 0.019 \citep{sch98}, which we convert assuming the standard galactic curve A$_\text{V}$ $=$ 3.1 E(B$-$V). Within the uncertainties, this internal extinction is in agreement with A$_\text{V, i}$ $=$ 0.40 as estimated by \citet{rei10} by fitting the spectrum with Starburst99 models. 

\subsection{\label{sec-nebular} Electron Temperatures, Density, and Pressure}    

In order to probe the pressure as well as measure oxygen abundance as a proxy for metallicity, we estimate the electron density and temperatures in S26. These physical conditions in the ionized gas of S26 are determined through ratios of line fluxes using the five-level atom model \citep{der87}  with the NEBULAR package in IRAF. The electron density is estimated using the SII line ratio 6716$\lambda$/6731$\lambda$ and the S$^{+}$ electron temperature given by the line ratio (6716$\lambda$+6731$\lambda$)/4076$\lambda$. The electron density is estimated to be n$_e$(SII)$\approx$75 cm$^{-3}$, although it may vary by as much as a factor of four due to propagation of flux uncertainties also impacting the temperature. The S$^{+}$ electron temperature is T(SII) $=$  6500 $\pm$ 800 K. The O$^{+}$ electron temperature is determined by T(OII) $=$ T(NII) \citep{izo94}, which results from the HII photoionization models of \citet{sta90}, and is measured to be T(NII) $=$ 13500  $\pm$ 1400 K using the [NII] ratio (6548$\lambda$+6584$\lambda$)/5755$\lambda$.  The O$^{++}$ electron temperature is measured to be T(OIII) $=$ 9400 $\pm$ 500 K using the [OIII] ratio (4959$\lambda$+5007$\lambda$)/4363$\lambda$. Using the estimated density of 75 cm$^{-3}$ then implies a pressure of P/k $=$ 7.5 $\times10^5$ cm$^{-3}$ K for S26. The temperatures, density, and pressure estimated for S26 fit the observed range of typical HII regions in disk galaxies and ELC regions in the Antennae Galaxy \citep{gg07,hh09}. We assume the standard HII temperature of 10$^4$ K as a representative single temperature throughout this work. 

\subsection{\label{sec-metallicity} Metallicity via Oxygen Abundance}

We derive the oxygen abundances using the standard T$_e$ method with two distinct temperature zones in the photoionized HII region, as in \citet{izo94, izo97}, using the O$^{+}$ and O$^{++}$ electron temperatures are explained above. The total oxygen abundance is derived by O/H = O$^{+}$/H$^{+}$+O$^{++}$/H$^{+}$: O$^{+}$ ionic abundance is measured with the summed flux of the doublet 7319/7330$\lambda$, and O$^{2+}$ ionic abundance is determined using ionic abundances for the lines 4363$\lambda$, 4959$\lambda$, and 5007$\lambda$. We measure a total oxygen abundance of 12 + log(O/H) = 8.3 $\pm$ 0.2, which is in agreement with checks using empirical relations from \citet{izo06} that were computed with new photoionization models. By assuming a simple scaling relation and the solar metallicity value 12+log(O/H)$_{\sun}$ $=$ 8.69 \citep{asp09}, the oxygen abundance converts roughly to a metallicity of Z$\sim$0.006  ($=$0.4  Z$_{\sun}$), similar to the LMC. The measured oxygen abundance of S26 is in excellent agreement with the value found by \citet{leq79} of 12+log(O/H) = 8.38. 

\subsection{\label{sec-age} Age}

\begin{figure}[t]
\includegraphics*[width=9cm,angle=0]{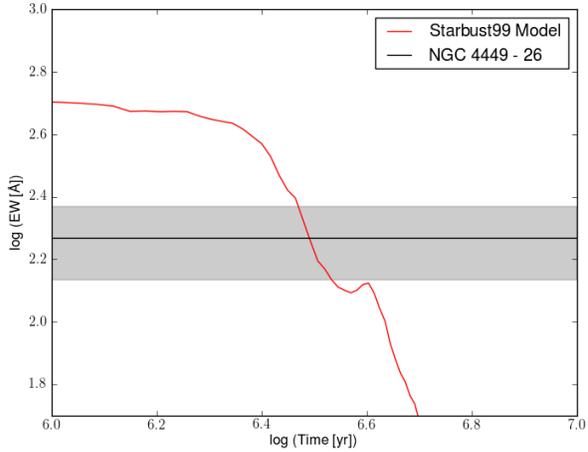}
\caption{\label{fig-sb99:age} Starburst99 predictions for the cluster age versus the equivalent width of the H$\beta$ (4861 \AA) line (see Section \ref{sec-sb99}). The observed EW of H$\beta$ for the massive star cluster S26 in NGC 4449 (solid line with uncertainties shown as a shadowed region) implies an age of 3.1 $\pm$ 0.3 Myr. Color version available online.}
\end{figure}

Ages are instrumental to characterizing the timescales of the early evolutionary stages of massive star clusters, as well as necessary for comparisons to predictions and other observed objects. The age of S26 is estimated by comparing the equivalent width of H$\beta$ to the evolutionary synthesis models of Starburst99 \citep{lei99}, using the measured equivalent width of H$\beta$ of $\approx$185 \AA, and adopting a metallicity of Z=0.008 (see details provided in Section \ref{sec-sb99}).  As shown in Figure \ref{fig-sb99:age}, we derive an age of 3.1 $\pm$ 0.3 Myr for S26, consistent with \citet{rei10}.

\section{\label{sec-pops} Wolf-Rayet Features and the Massive Star Populations}

The first hint of the important evolutionary phase of S26 was the discovery of an optical feature due to Wolf-Rayet stars \citep{rei10}. WRs are spectroscopically identifiable via unique emission lines and are divided into subtypes of nitrogen-rich WNs, carbon-rich WCs, and oxygen-rich WOs; the presence of several WR stars will produce broad features in the integrated optical spectrum known as  WR ``bumps'' (Fig. \ref{fig-wrspec}). S26 clearly displays broad WR features of both WN and WC populations, namely the ``blue bump'' near 4650 \AA, consisting of a blend of [NIII] 4640 \AA, [CIII] 4650 \AA, and HeII at 4686 \AA, and the ``red bump'' due to [CIV] at 5808 \AA. The emission at 4686 \AA\ of S26 does not display a clearly distinct nebular component (Fig. \ref{fig-wrspec} ) often seen in the integrated spectra of WR galaxies e.g. \citet{con91}. However, at relatively low metallicities, such as S26, the stellar component is thought to dominate \citep{sv98}. The expected nebular emission for a population of 3 - 3.5 Myr should be log($\frac{\text{I(nebular HeII [4686 \AA])}} {\text{I(H$\beta$)}}$)$\sim$-2 according to \citet{sv98}. Thus, the expected nebular emission is too low to be significantly detected or distinguished from the stellar emission at 4686 \AA\ with this dataset, and thus the lack of resolved nebular emission is not surprising. Other WR features that could further classify the WR population would be similarly too faint to detect.

\subsection{\label{sec-wrstars} Determining the Number of WR stars}

The luminosity of a WR feature can be used to constrain the WR population and estimating the WR star populations in an extragalactic source has become fairly standardized \citep[e.g.][]{sv98,gus00}.  The number of WR stars can be estimated as $N_{\text{WR}}=L_{\text{WR}}/L_{\text{o,WR}}$, with $L_{\text{o,WR}}$ as a typical single WR star producing the feature and $L_{\text{WR}}$ as the observed luminosity of that feature from the source. The subtypes can be separately analyzed with their respective WR lines. Many uncertainties result from this method due to the inherent range of WR line fluxes emitted by individual stars, averaging over WR subtypes, and the currently poor understanding of the impact of environment on WR line fluxes. Nonetheless, it is useful to estimate the massive star populations within a cluster, as long as the range of uncertainties are kept in mind while interpreting the results.

In order to constrain the WR subtype populations in S26,  we adopt the method of \citet{gus00}. The red bump at 5808 \AA\ is thought to be produced only by WCE stars and can be compared to representative WC4 stars. The number of WC stars can be approximated as $N_{\text{WC}}=L(5808 \text{ \AA})/L_{\text{WC4}}(5808 \text{ \AA})$, assuming a typical luminosity of a single WC4 star at 5808 \AA\ is 3.0$\times$10$^{36}$ ergs s$^{-1}$ as measured in the LMC \citep{sv98}.

Determining the number of WN stars is less straight forward because the blue bump includes contributions from both WN and WC stars. The relative contribution of WC stars to the blue bump feature can be estimated from the red bump, unique to the WCs, and is described by the coefficient $k=L_{\text{WC4}} (4650  \text{ \AA})/ L_{\text{WC4}} (5808 \text{ \AA})$ \citep{gus00}. We adopt a value of k$=$1.71 $\pm$ 0.53 \citep{sv98,gus00}, although uncertainties are large due to variations in relative line fluxes \citep{sv98}. After subtracting the estimated WC contribution from the total measured flux in the blue bump, the number of WN stars is found by comparing the remaining emission to a typical WN star.  We assume a typical WN of a WNL (WN7) star to be with a luminosity of 2.0 $\times$ 10$^{36}$ erg s$^{-1}$ in the blue bump \citep[4650 $+$ 4686 \AA; ][]{gus00} as in the LMC. 

In total, we have estimated 20 $\pm$ 14 Wolf-Rayet stars in S26, comprised of roughly 4 $\pm$ 3 WC and 16 $\pm$ 13 WN stars. The results can be found in Table \ref{table-pops}. The uncertainties are estimated from assuming a typical flux, where we use the observed flux range in the LMC by \citet{sv98}, as well as from flux measurement uncertainties in Table \ref{table-flux}. Due to the large range of observed variation of fluxes, the uncertainties are correspondingly, and unavoidably, large. Another potential uncertainty that has not been accounted for arises from the weak HeII line emission from the most massive O-stars (Of-stars), which may result in an overestimate of WN stars. However, this effect is expected to be small at the metallicity and age of S26,  roughly Of/O$\sim$0.15, which with the weak HeII emission (10\% of a WN star) would contribute a total of $\sim$2 WN stars \citep{sv98}. Additionally the ratio of WC to WN stars appears normal, see Section \ref{sec-wrpop}, and thus Of-star contamination is likely not important.

\begin{deluxetable}{ll}
\tabletypesize{\scriptsize}
\tablewidth{0pc}
\tablecaption{\label{table-pops}Massive Star Populations in S26} 
\tablehead{
  \colhead{Type}  &
  \colhead{Number}  
}
\startdata

O		&  219	$\pm$ 15 \\
WNL     &  15		$\pm$ 12	\\
WCE		&  3		$\pm$ 3	\\
WR$_{\text{total}}$	&  18	 $\pm$ 13	\\
\\
\hline
\\
WR/O      	&  0.084	$\pm$ 0.058 \\
WC/WN   	&  0.23 $\pm$ 0.26	\\
\enddata
\end{deluxetable}

\subsection{\label{sec-ostars} Determining the Number of O-stars}

Identifying the total massive star population in S26 is important for understanding the feedback processes that are altering the cluster. In addition to the massive WR stars that are driving strong winds and hard radiation throughout the cluster, the O-star populations will also contribute ionizing radiation and additional (albeit weaker) winds. The number of O-stars can be estimated by determining the population needed to produce the observed ionizing photons Q$_{\text{o}}$, after subtracting off the Wolf-Rayet contributions. The ionizing flux seen at optical wavelengths is simply estimated through empirical relations from \citet{sv98}.  Using H$\beta$ at 4861 \AA, the ionizing flux is Q$_{\text{o}}$$\sim$170$\times 10^{49}$ photons s$^{-1}$. We can thus estimate the number of O-stars by assuming $$N_{\text{O}}=(Q_{\text{o}} - N_{\text{WR}} Q_{\text{o,WR}})/(\eta_o Q_{\text{o,O7V}})$$ \citep{gus00}. We include the parameter $\eta_{\text{o}}$ as the ratio of O7V stars to all O-stars to account for different O-star subtypes occurring within IMF (as subtypes produce different ionizing photon fluxes). At an age of 3.1 Myr and Salpeter IMF, we find  $\eta_{\text{o}}$$\sim$1.2 for the cluster S26 \citep{sv98}, near the peak value resulting from a WR rich phase.  A typical ionizing photon flux for an O-star is $Q_{\text{o,O7V}}$ is taken to be 10$^{48.75}$ s$^{-1}$ from an O7V star \citep{mart05} and 10$^{49}$ s$^{-1}$ for a WR star \citep{gus00,sch99}. We estimate O-star population uncertainties by accounting for the measured flux uncertainty in S26, estimated uncertainties on $\eta_{\text{o}}$, and uncertainties for the subtracted WR populations. With a WR population of  $N_{\text{WR}}=$ 18 from above, we find there are approximately 219 $\pm$ 15 total O-stars harbored in the massive star cluster S26.

\section{\label{sec-radiothermal} The Thermal Radio Component of S26}

The radio spectral indices of S26 indicate that the source hosts mixed thermal and non-thermal contributions; and as shown in Figure \ref{fig-hst}, the radio continuum contours additionally change from a resolved, irregular shape at 3$\sigma$ to an unresolved, compact peak at 6$\sigma$. Despite the messy nature of S26, it is necessary to identify the radio emission produced by the HII region itself in S26. We decompose the observed radio emission to the published fluxes at 1.3 cm, 3.6 cm, and 6.0 cm from \citet{rei08} and assume the thermal emission follows $F_{\nu} \propto \nu ^{-0.1}$ and non-thermal emission as $F_{\nu} \propto \nu ^{-0.7}$ \citep[e.g.][]{baar77}. The thermal emission is, however, dependent on both the size and density of the emitting region as $S_{\nu, {\text{thermal}}} \propto 2kT\nu^2 \tau_\nu /c^2 $, and the models are thus under-constrained. To better constrain the fit, we can adopt the approximate size of the 3.6 cm radio continuum emission region. The 3$\sigma$ contour suggests a  radius of $\sim$50 pc -- however a smaller size would be derived if only the unresolved peak emission was used \citep[][also seen in Figure \ref{fig-hst}]{rei08}. 

\begin{figure}[t]
\includegraphics*[width=9cm,angle=0]{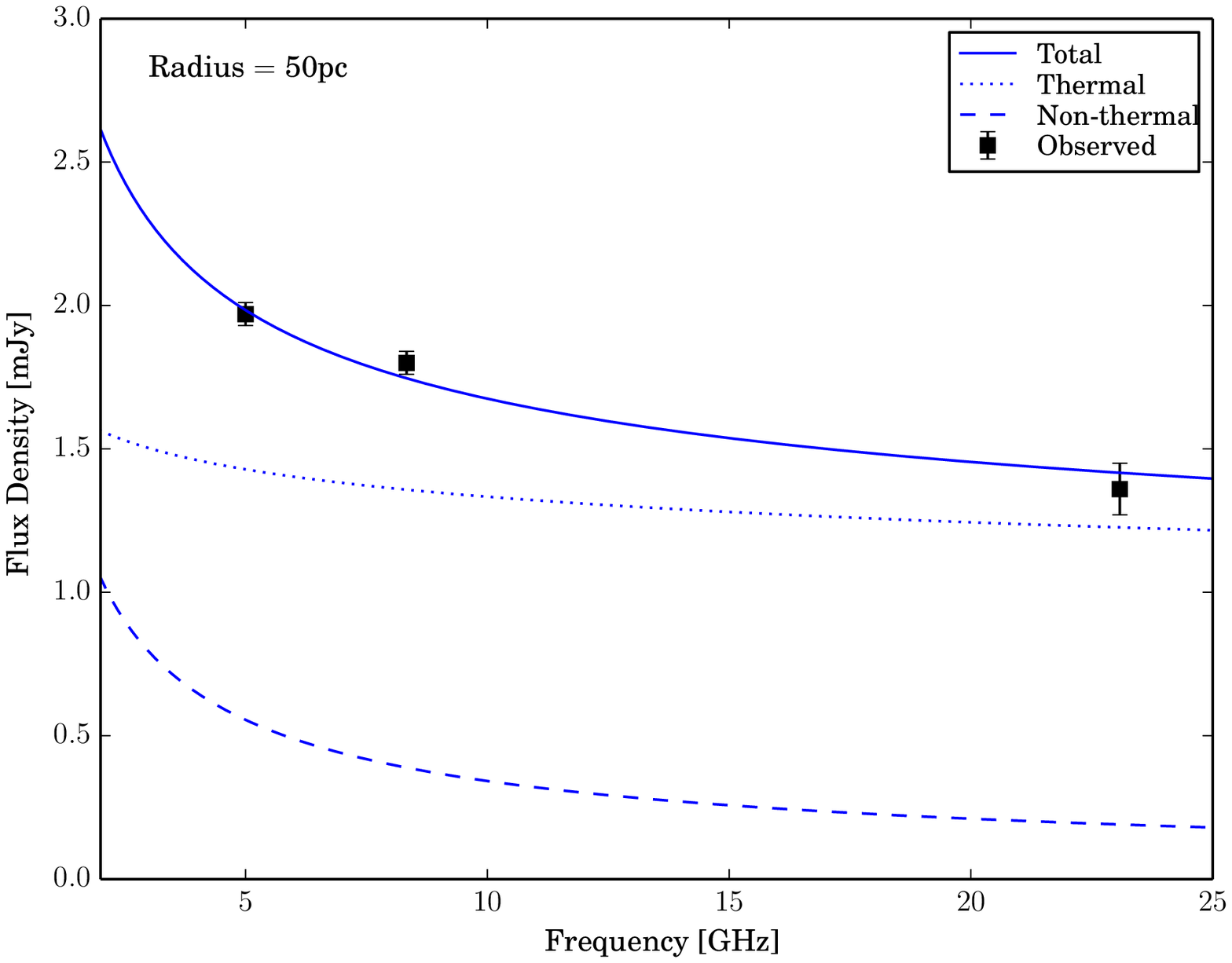}
\includegraphics*[width=9cm,angle=0]{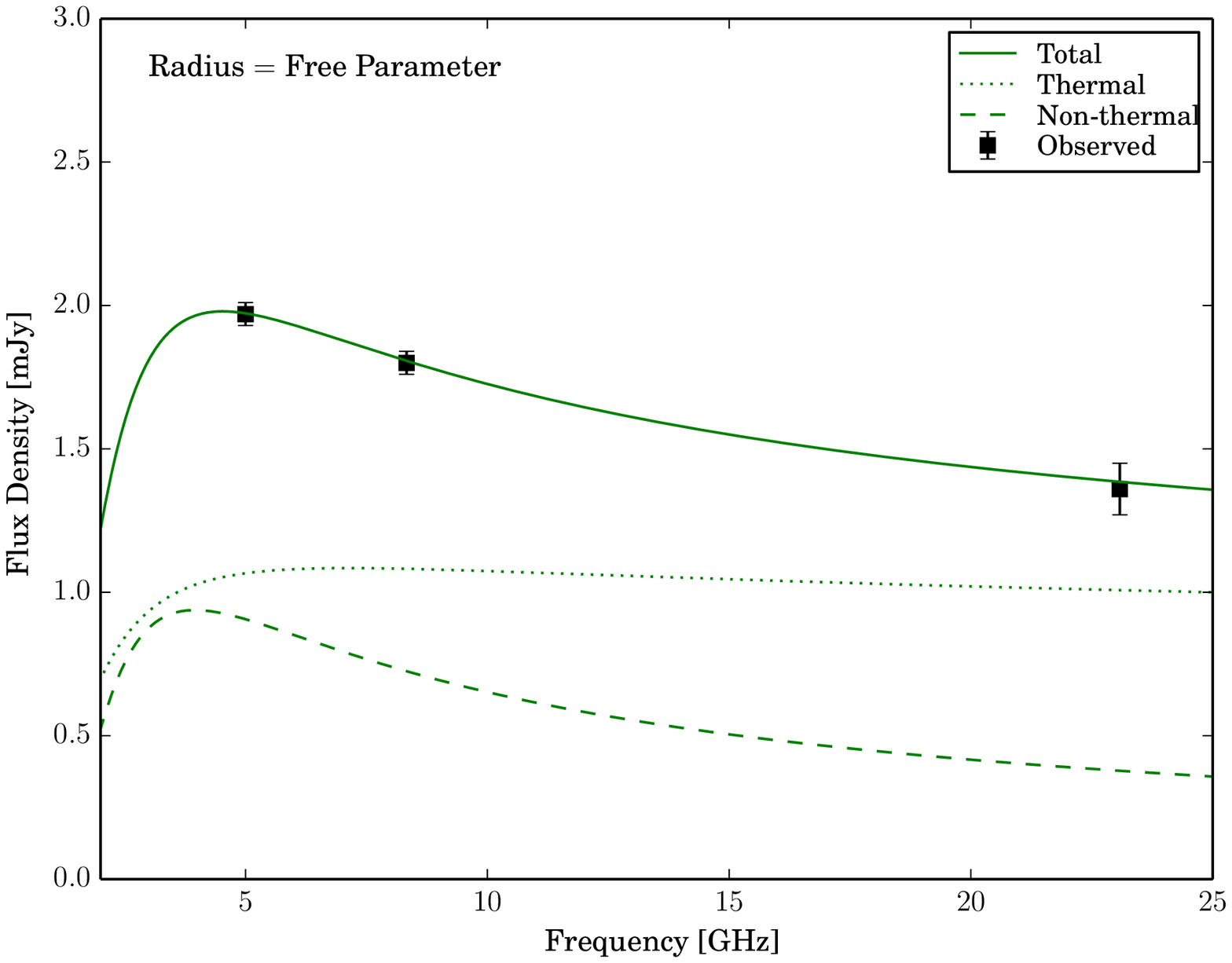}
\caption{\label{fig-radiothermal} Identifying the thermal radio component in the radio SED of S26, as observed by \citet{rei08}, with a $\chi^2$ fit to the observed fluxes of the expected emission assuming thermal emission goes as $F_{\nu} \propto \nu ^{-0.1}$ while non-thermal emission as $F_{\nu} \propto \nu ^{-0.7}$. Left: Constraining the fit such that the radius is set to 50 pc, as observed by the irregular 3$\sigma$ contour of the radio continuum, we find that the flux has non-thermal contributions of 14\% at 1.3 cm. Right: Allowing the radius to be a free parameter (although under-constraining the fit) results in an increased non-thermal contribution of 28\% at 1.3 cm from a region of radius 2.3 pc. Color version available online.}
\end{figure}

We find the model with the lowest $\chi^2$ by both setting the radius and leaving it as a free parameter, as shown in Figure \ref{fig-radiothermal}. By imposing a radius of the adopted value of 50 pc, we find the density is $n_e =$ 21.5 cm$^{-3}$ and the non-thermal contributions at 1.3 cm are 14\% $\pm$ 15\%, the band least contaminated by non-thermal emission. If the radius is unconstrained, we find the thermal radio emission results from a region of radius of r$=$ 2.3 pc, a density of $n_e =$ 2.0$\times 10^3$ cm$^{-3}$, and non-thermal contributions at 1.3 cm of 28\% $\pm$ 25\%. The fit is under-constrained for both models ($\chi^2_{\text{red}}$$\sim$0.5 and $\sim$0.1 for a set and free radius) and the equality of the fits cannot discriminate between the two conditions and rather express both as possibilities. 

Thus, the radio data is consistent with both a large, low-density HII region typical of Giant HII regions (GHRs) and a very dense, compact region typical of UDHIIs (or, more likely, a combination of the two, although the data do not permit a more sophisticated model).  S26 clearly has associated extended thermal radio emission, thus there may be a low filling factor of small dense regions or a single very dense region within a larger, less dense S26 HII region. This type of scenario has been observed in Mrk 996, where the extremely dense (n$_{e}$$\sim$10$^6$ cm$^{-3}$) nucleus is surrounded by a less dense ($\sim$10$^2$ cm$^{-3}$) star forming region, along with WR stars \citep{tel14}. We find the flux at 1.3 cm of  S$_{\nu \text{,1.3 cm}}$$\sim$1360 $\mu$Jy is roughly representative of the thermal radio emission. While S26 is not solely a thermal source, decomposition shows the flux at 1.3 cm can be considered primarily thermal.

While the majority of the thermal radio emission may be emitted by an HII region surrounding the cluster, the Wolf-Rayet populations will also contribute to the observed radio flux. Strong WR winds are known to produce thermal free-free radio emission, approximated by the scaling relation $$S_\nu \propto \frac{\dot{M}}{v_{\infty}^{4/3}} \frac{\nu^{0.6}}{D^2}$$ \citep{cro07}. Using this relation, we can estimate the expected radio flux density contribution from the Wolf-Rayet stars in S26 to identify whether the HII region or the WR winds are the likely dominant emitter. Accounting for metallicity, we can estimate the expected radio flux density by scaling from galactic WR stars observed with the Very Large Array \citep{cap04}. Exactly how mass-loss rates depend on environment is not well known, although mass-loss rates at low metallicities are observed to be lower than higher metallicity environments as \.{M}$_{\text{WN}}$ $\propto$ Z$^{0.7}$ and \.{M}$_{\text{WC}}$ $\propto$ Z$^{0.5}$ \citep{cro06}. Clumping in the winds can additionally reduce mass loss rates by factors of two \citep{cro07}.  Taking a generic WC star for example, WR 5 (WC6 star) with S$_\nu$ $=$ 0.20 mJy at 3.6 cm \citep{cap04} would produce S$_{\text{3.6 cm}}$$\sim$5.2 $\times$ 10$^{-8}$ mJy at the distance and metallicity of S26 in NGC 4449. Similarly, the radio flux density of a WN star could produce S$_{\text{3.6 cm}}$$\sim$5.6 $\times$ 10$^{-7}$ mJy (e.g. the WN 7 star WR 100). Summing the total population of WR stars in S26 and scaling to 1.3 cm, we might expect a flux density of around S$_{\nu \text{,1.3 cm}}$  9 $\times$10$^{-6}$ mJy due to the WR winds, which compared to the observed flux of S$_{\nu \text{,1.3 cm}}$$\sim$1.360 mJy, is negligible. Therefore, the thermal radio emission at 1.3 cm resulting from the WR winds is 1:$10^{-5}$ to the thermal radio emission observed in S26 and the majority must be resulting from free-free emission from the HII region ionized by the stars.

\section{\label{sec-dust} Dissecting the Dust}

We utilize archival data of the galaxy NGC 4449 from near-infrared to far-infrared wavelengths to construct the spectral energy distribution (SED) of the massive star cluster S26. The infrared emission from S26 is clearly one of the dominating sources in its host galaxy, shown in Figure \ref{fig-irview},  and at the longest wavelength (500 $\mu$m) is comparable in brightness to the nuclear emission. 

\subsection{\label{sec-photometry} Photometry}

Embedding material surrounding S26 is evident throughout the rich archival dataset of NGC 4449 and can be characterized through photometry and the construction of an SED. An SED provides key information on the heated dust properties and is crucial in describing the evolutionary phase. However, the wavelength coverage from a few to hundreds of $\mu$m, along with source complexity as well as corresponding worsening resolution, necessitates careful treatment of the different datasets to preserve the shape of the SED. 

We perform photometry using two apertures to evaluate both the resolved emission and the total emission, which additionally will provide insight into any radial trends; the apertures are overlaid in Figure \ref{fig-irview} and explained below. Aperture photometry is performed using the IDL procedure SURPHOT \citep{rei08}. The uncertainties are dominated by background subtraction, and these are estimated empirically by calculating the standard deviation of the fluxes measured using different backgrounds. 

Using native PSFs, we use a small extraction aperture for images where S26 is resolved from nearby sources (images centered at 100 $\mu$m or less). We adopt a small circular aperture with a radius of 5'' ($\sim$100 pc; see Figure \ref{fig-irview}), which sufficiently excludes nearby sources while including the resolved emission from S26.  We follow the IRAC Instrument Handbook for Calibration of Extended Sources \url{http://irsa.ipac.caltech.edu/data/SPITZER/docs/irac/iracinstrumenthandbook/29/#_Toc296497401} as well as the procedure in \citet{rr13} and the encircled energies in \citet{bal14} for extracting fluxes from PACS data, in agreement with \url{http://herschel.esac.esa.int/Docs/HerschelUG/HUG2web_BA_PACSExtendedSourcePhotometry.pdf}

The total emission is measured with a large aperture over the complete dataset. However, resolution matching is necessary due to the large change in PSF with increasing wavelength for the IR datasets. Therefore, we construct the total emission SED using a large aperture extraction region (a 15''$\sim$300 pc radius) on images convolved to a common resolution  (SPIRE 500 $\mu$m  with a PSF of 36'')  using convolution kernels provided by \cite{ani11}. The aperture correction for images convolved to the SPIRE 500 $\mu$m PSF and extracted with the large aperture are found by applying the same photometry procedure to the corresponding PSF  and the same extraction apertures, as the region size is sufficiently small enough to approximate a point source at this PSF.

\subsection{\label{sec-sed} Spectral Energy Distribution of S26}

\begin{figure}[t]
\includegraphics*[width=\textwidth,angle=0]{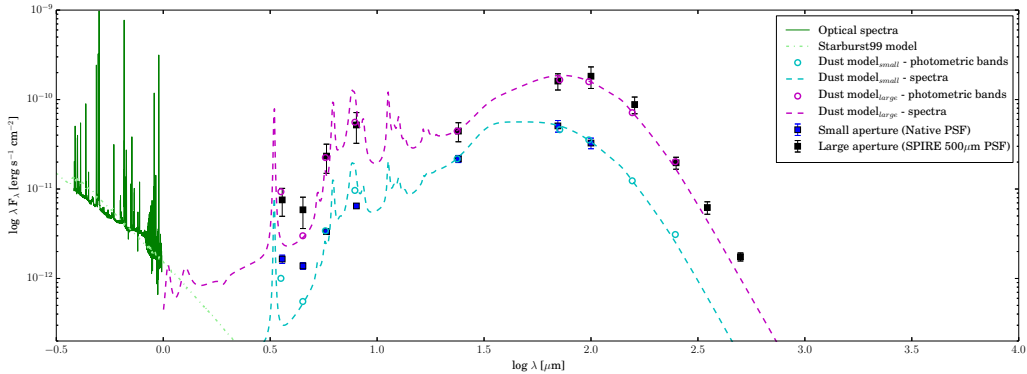}
\caption{\label{fig-sed} An SED of the optical to far-IR emission from the massive star cluster S26 with dust and grain models over plotted for comparison. Optical spectra (green line) is from \citet{rei10}, and a Starburst99 model (dotted green line) clearly demonstrates an abundance of IR emission to the expected stellar emission. The IR photometry and dust and grain model fitting utilize two apertures, explained in further detail in Section \ref{sec-dust}: A small 5'' aperture (blue squares used to extract emission from images with native PSFs, and a large 15'' aperture (black squares) used to extract emission from images convolved to the resolution of SPIRE 500 $\mu$m. The best fit dust and grain model from \citet{draine07} are plotted for the corresponding aperture (large or small), where the dotted line shows the model spectra and the empty circular dots show the photometric fluxes. The SED clearly shows strong PAH features and large amounts of dust surrounding S26. The change in PAH emission in the IRAC bands at 5.8 $\mu$m and 8.0 $\mu$m and the shift in the peak IR emission can be seen between the two IR aperture SEDs, plausibly due to increased exciting radiation and destruction of PAHs towards the cluster center.}
\end{figure}

The spectral energy distribution (SED) of S26 is shown in Figure \ref{fig-sed}; the shape and peak result from large amounts of heated dust. For comparison, we plot the expected stellar emission from the Starburst99 model (see Section \ref{sec-sb99})  of a cluster of similar mass undergoing an instantaneous starburst. It is clear that the infrared emission cannot be produced solely by stellar sources. Instead, this emission is produced from a combination of heated dust partially embedding the cluster and Polycyclic Aromatic Hydrocarbon grain (PAHs) emission in the photodissociation region (PDR) surrounding the ionizing stars of the cluster. Strong evidence for PAH emission is observed as an excess from the IRAC 5.8 $\mu$m and 8.0 $\mu$m - viewed as a bump between the IRAC 4.5 $\mu$m  and MIPS 24 $\mu$m points. The infrared emission is extracted using two different methods due to the drastic change in resolution at increasing wavelengths, as explained above. Not surprisingly, the SED extracted from the large aperture results in more emission than the SED resulting from the smaller aperture.

\subsection{\label{sec-dustmodel} The Best Fit Dust and Grain Model}

SEDs can be quantitatively analyzed to describe the emitting material: to characterize the dust, PAH components, and exciting radiation, we compare dust and grain models of a gas and dust mixture heated by a distribution of starlight intensities from \citet{draine07} to the infrared SED of S26. The starlight intensity is given by a scaling factor U such that the energy density per unit frequency is $u_\nu=Uu^{\text{MMP83}}_\nu$, where the interstellar radiation field $u^{\text{MMP83}}_\nu$ is from \citet{mat83}. The fraction of the dust mass that is exposed to a distribution of intensities from $U_{\text{min}}$ to $U_{\text{max}}$ is described by the parameter $\gamma$; therefore (1-$\gamma$) is the fraction exposed to $U_{\text{min}}$. The fraction of the total dust mass that is in PAH particles is $q_{PAH}$ and expected to be low at the low metallicity of NGC 4449.

The best fit models to the S26 SEDs are found via a $\chi^2$ test between the observed photometric data for S26 and a model's photometric flux at corresponding bands. The characteristics of the best fit models are listed in Table \ref{table-sed} and the fits are shown in Figure \ref{fig-sed}. The best fit models suggest that the exciting radiation in S26 is strong, with a maximum starlight intensity of U$_{\text{max}}=$10$^6$, which is similar to starbursting galaxies. The minimum exciting intensity may increase towards the cluster center, as U$_{\text{min}}=$12.0 for the best fit model to the large aperture SED and U$_{\text{min}}=$25.0 for the best fit model to the small aperture SED. Compared to a global estimate of 2\% PAH emission in NGC 4449 \citep{kar13}, the best fit models may additionally indicate this changing environment towards the center of S26. The large aperture SED (extracted from a radius of $\approx$300 pc) best matches a model with high PAH emission (3.19\%) and nearer to the center of the cluster, the small aperture SED (extracted from a radius of $\approx$100 pc) best matches a model where the PAH emission is reduced to 1.12\%. Thus, the dust grains may be being destroyed due to a higher radiation field.

\begin{deluxetable}{lll}
\tabletypesize{\scriptsize}
\tablewidth{0pc}
\tablecaption{\label{table-sed} SED of S26: Photometry and Model Parameters} 
\tablehead{
\colhead{}  &

 \colhead{Small Aperture}   &
\colhead{Large Aperture} 
}
\startdata
\\
Band				&	Flux Density (mJy)	& Flux Density (mJy)		\\
\hdashline
IRAC 3.6 $\mu$m	&	1.98 (0.21) 	&	9.1 (3.1)	\\
IRAC 4.5 $\mu$m	&	2.07 (0.18) 	&	8.8 (3.4)	\\
IRAC 5.8 $\mu$m	&	6.46 (0.42)	&	45 (16) 	\\
IRAC 8.0 $\mu$m	&	17.24 (0.93)	&	139 (52)	\\
MIPS 24 $\mu$m	&	174 (17)		&	355 (88)	\\
PACS 70 $\mu$m	&	1180 (180)	&	3760 (770)	\\
PACS 100 $\mu$m	&	1090  (150)	&	6100 (1600)	\\
PACS 170 $\mu$m	&	...	&	4690 (990)	\\
SPIRE 250 $\mu$m	&	...	&      1640 (250) \\
SPIRE 350 $\mu$m	&	...	&	720 (120)	\\
SPIRE 500 $\mu$m	&	...	&	290 (30) 	\\

\\
\hline
\\
$U_{\text{min}}$		&	25.0			&  12.0		\\
$U_{\text{max}}$		&	$10^{6}$		&  $10^{6}$		\\
$\gamma$    			&	0.08			&  0.03		\\
$q_{\text{PAH}}$		&	1.12\%		&  3.19\%	\\

\enddata
\tablenotetext{a} {
Notes: The upper part of the table presents the flux densities extracted from the corresponding aperture. The lower part of the table presents the parameters determined by the best fit models from \citet{draine07}, which describe a gas and dust mixture heated by a distribution of starlight intensities, U, where the local starlight intensity is U=1. The fraction of the dust mass exposed to $U_{\text{min}}$ is 1-$\gamma$. The emission due to the PAH particles is given as $q_{\text{PAH}}$.
}
\end{deluxetable}

\subsection{\label{sec-dustmass} Dust Mass and Star Formation Efficiency}

Dust is an integral component to the natal cluster environment. Understanding the dust mass that is surrounding a region is important in determining the evolutionary stage, providing comparisons of similar regions to formulate an observational picture, and for extrapolating to estimate the gas mass. The ratio of the stellar to gas mass serves an indicator for the degree of removal of natal material. Utilizing results of the dust model fitting and the observed photometric fluxes, the dust mass surrounding the cluster S26 can be estimated. Following the method of \citet{draine07}, we find a total dust mass of 2.3 $\times 10^5 \text{ M}_{\sun}$ as measured with the large aperture (with uncertainties of 15\% due to flux uncertainties alone). Due to the resolution at the long wavelengths necessary for the dust mass estimate, we only have a measurement over this spatial scale, which is $\sim$300 pc.

This total dust mass, which is sampled over a large region, is roughly three times the stellar mass estimated by \citet{rei10} of 6.5$\times 10^4  \text{ M}_{\sun}$, which corresponds to a  region of r$=$3''.3. However, we can scale the stellar mass to the same large 15'' aperture used for the dust mass estimate. We determine this scalefactor by comparing the flux measured in an archival {\em Hubble Space Telescope} I-band image extracted with the r$=$15'' circle (as above) to the value published in \citet{rei10}. We find the scaled stellar mass in S26 over the large aperture is 3.0 $\times 10^5  \text{ M}_{\sun}$. To estimate the total gas mass surrounding the cluster as well, we assume a dust-to-gas ratio suggested by the best fit model from \citet{draine07} of 1/130, which gives a gas mass of 3.0 $\times 10^7 \text{ M}_{\sun}$.  The uncertainties are at least 50\% based on the range of acceptable models and adopting a dust-to-gas ratio. For comparison to the host galaxy, modeling by \citet{kar13} found a global value of dust-to-gas ratio equal to 1/190 in NGC 4449. Regions of higher star formation events such as in LIRGs show average values of the gas to dust ratios of 120 $\pm$ 28 \citep[Luminous Infrared Galaxies  - LIRGS;][]{wil08}, in excellent agreement with the model value found for S26. By adopting the above total gas mass, the star formation efficiency (SFE) of the entire S26 region can be estimated.  

A gas mass this large implies a low {\it global} SFE of 1\% for the region of r$\sim$300 pc surrounding S26. Typical values for bound clusters are observed to be SFE$\approx$20-50\% \citep{ash01,kro01}. However, low SFEs can be observed, such as 5-10\% over entire molecular clouds \citep{wil97}, and the size scale over which SFEs are measured can have a major role. Additionally, the extremely low inferred SFE of S26 may result from several factors that we cannot discriminate between. The extraction region for the large aperture SED is roughly 35 times the area used to estimate the stellar mass of the cluster, and thus may be including dust and gas beyond the S26 structure. Alternatively, we cannot rule out if S26 is done forming stars or if the massive stars are disrupting or inhibiting further star formation that might be occurring. 

\section{\label{sec-discussion} Discussion}

It is evident that the massive star cluster S26 in NGC 4449 is emerging from its natal cocoon, as seen through multi-wavelength evidence: 1. thermal radio emission from ionized gas indicative of a young massive star cluster,  2. unique optical features produced by the evolved massive Wolf-Rayet stars producing large amounts of mechanical stellar feedback, and 3. strong infrared and PAH emission from a PDR and heated dust surrounding the cluster. Here we  investigate physical conditions pertaining to the massive stars within the cluster, which may be driving or contributing to this cluster's emergence from its birth material. 

\subsection{\label{sec-popsincontext}  Massive Star Populations of S26 in Context}

\subsubsection{\label{sec-sb99} Estimates through Starburst99}

\begin{figure}[t]
\includegraphics*[width=9cm,angle=0]{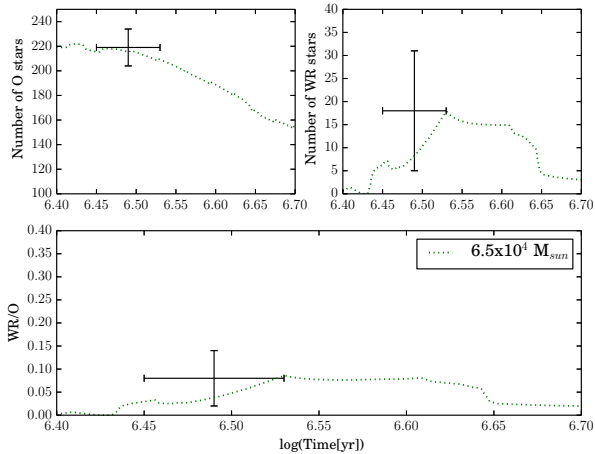}
\caption{\label{fig-sb99:wro} 
The estimated age versus the observed WR and O-star populations, and WR/O population ratio, of the cluster S26 (data points). The evolutionary synthesis model of Starburst99 \citep{lei99}, which utilizes a Kroupa IMF with upper and lower limits of 0.1 - 120 M$_{\sun}$ , is plotted (dotted lines) and further discussed in \ref{sec-sb99}. Color version available online.}

\end{figure}

Models and population comparisons may show evolution timescales, metallicity or environmental effects, or deviations from the expected IMF, and thus indicate processes that are affecting the star formation. Using Starburst99 v7.0.0 \citep{sb99}, the starburst properties of S26 are modeled by adopting an input metallicity of Z = 0.008 (see Section \ref{sec-nebular}) and simulating an instantaneous burst of star formation. The Geneva evolutionary tracks with high mass loss and Pauldrach/Hillier atmospheres \citep{smi02} are used with a Kroupa IMF. The Starburst99 models are used to determine the age (Section \ref{sec-age}) and provide an evolutionary comparison for theoretical massive star populations, as shown in Figure \ref{fig-sb99:wro}.   To produce comparable massive star populations to the observed data, we scale the initial cluster mass to 6.5$\times$10$^4$ M$_{\sun}$ \citep{rei10} and use an upper mass limit in the IMF of 0.1 - 120 M$_{\sun}$. The observed WR/O population ratio in the massive star cluster S26 is consistent with these predictions, but is slightly higher than expected, as shown in Figure \ref{fig-sb99:wro}. If the age were a little bit larger, this discrepancy would go away. In general, we expect fewer Wolf-Rayet stars at low metallicities like S26. 

\subsubsection{\label{sec-wrpop} Wolf-Rayet Population Trends with Metallicity}

\begin{figure}[t]
\includegraphics*[width=1.0\textwidth,]{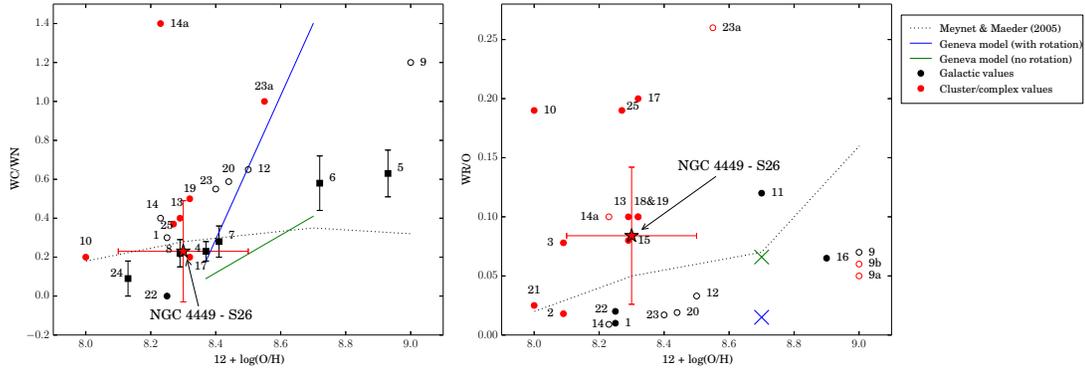}
\caption{\label{fig-popratios} 
A compilation of observed population ratios in nearby galaxies compared to different predictions. 
Star points show the observed ratios in the massive star cluster S26 in NGC 4449. Lines show model predictions of \citep {mm05}   (dotted; including stellar rotation) and computed results for new models from the Geneva group (solid) for z$=$0.006 (Neugent et al. 2012) and z$=$0.014 (Georgy et al. 2012) in blue and green. The symbols roughly indicate the method of determining the populations:  squares show individually, spectroscopically resolved populations and circles show ratios similarly measured to this work (open are corrected for completeness). Color version available online. References:
1-IC 10 \citep{mas02}, 
2,3-IC 4662 A1,A2  \citep{cro09}, 
4-LMC\citep{neu12}, 
5-M 31 \citep{neu12},
6,7,8-M 33 inner, middle, outer \citep{neu12},
9/9a/9b-M 83/M 83-74/M83 - 31 \citep{had08},
10-Mrk 996 nuclear \citep{tel14},
11-MW \citep{geo12},
12-NGC 300 \citep{bib10},
13-NGC 1140 \citep{mol07},
14-NGC 1313 \citep{hc07}, 
15-NGC 1569 SSCA \citep{gd97} (under S26 in WR/O: red filled circle),
16-NGC 3049 \citep{gd02},
17,18,19-NGC 3125 A1,A2,B \citep{had06},
20-NGC 5068 \citep{bib12},
21-NGC 5253 B \citep{sid04},
22-NGC 6822 \citep{mj98},
23/23a-NGC 7793/NGC 7793 R34 \citep{bib10},
24-SMC \citep{neu12},
25-Tol 89 \citep{sid06}.}
\end{figure}

It is necessary to additionally compare the massive star populations of S26 to observations of other regions and models that are in different environments. The number of Wolf-Rayet stars, specifically comparisons between the subtype of the WRs, is heavily dependent on the metallicity. The ratios of Wolf-Rayet populations WC/WN and WR/O stars in the local group are observed to roughly decrease with host galaxy metallicity \citep[e.g. ][]{mas96}. Figure \ref{fig-popratios} shows evolutionary predictions against observational data of nearby galaxies and S26 in NGC 4449. Evolutionary model predictions of the Geneva group are shown, including \citet{mm05}, for initially rotating single stars, and new results presented in \citet{neu12} for Z=0.006 and \citet{geo12} for Z=0.014, with and without including stellar rotation. We convert from theory predictions at each metallicity Z to 12+log(O/H) assuming a simple scaling relation and 12+log(O/H)$_{\sun}$ $=$ 8.69 \citep{asp09}. 

Similar plots displaying metallicity trends of Wolf-Rayet populations are shown by the surveys from \citet{bib10} as well as a very thorough sampling by \citet{neu11} and \citet{neu12}. Here, we compare the inferred stellar populations in S26 to a larger sample of other massive clusters and galaxies found in the literature (references given in the caption of Figure \ref{fig-popratios}). In several cases, observational datasets have been evaluated for completeness and/or individual stars have been spectroscopically confirmed as Wolf-Rayet stars \citep[e.g.][]{neu12,bib10}. 

The inferred subtype WC/WN ratio of S26 roughly agrees with predictions (Fig. \ref{fig-popratios}) in the low metallicity cases, especially in the cases that have been corrected for completeness. The observed WC/WN ratio in S26 in NGC 4449 does not appear unusual in comparison to other galaxies. 

In contrast to the relatively well-behaved WC/WN ratio, scatter is seen between the WR/O ratio observations and predictions at high WR/O ratios that has received attention in only a few cases. This is rather surprising, as recent comparisons of the observed WR/O ratio to predictions in the solar neighborhood were used to suggest that single star evolution may only account for 60\% of WRs \citep{geo12}. As Figure \ref{fig-popratios} shows, a few observed regions populate an area on the plot of low metallicity yet high WR/O ratios. These values often correspond to SSCs rather than values from integrated regions across galaxies (although the distinction becomes unclear for the smallest galaxies). For instance, points plotted for NGC 3125 and Tol 89 are known intense star-forming regions. There was debate over how high the intensity of the starburst and WR populations in NGC 3125 really are \citep{sch99,chan04,had06,wof14}, and in this case the high WR/O ratio was broadly discussed. Tol 89 is one of the brightest known GHRs and has been resolved into individual compact clusters, four of which contain WR clusters \citep{sid06}. These regions clearly display large WR/O ratios which might then suggest some sort of extreme star formation indicator. Most recently, the unusually high WR/O ratio in NGC 3125-A1 has be interpreted to suggest the upper mass limit of the IMF is $>$ 120 M$_{\sun}$ \citep{wof14}. While the massive star cluster S26 in NGC 4449 is observed in the middle of the observed WR/O ratios and within the uncertainties is consistent with predictions, S26 is currently amid regions of intense starburst nature.

\subsubsection{\label{sec-considerations} Additional Considerations}

Many assumptions are necessary to estimate massive star populations, each with caveats that could impact the inferred stellar content. First, the O-star and WR estimates would be incorrect if the contamination by Of-stars were underestimated, or if the value of the standard ionizing flux Q$_{\text{o}}$ for any of the subtypes were incorrect or changed. In fact, use of the ionizing flux Q$_{\text{o}}$ $=$ 10$^{48.75}$ from \citet{mart05} results in twice the O-star population than that of the canonical value of 10$^{49}$ from \citet{lei99}. Additionally, the age of S26 is important in estimating $\eta_o$ to correct from an O7V population to the complete O-star population and, even more so, for comparing to the predictions of Starburst99. Lastly some WR stars can form from binary systems \citep{geo12}, however we compare to models which include only single star evolution tracks as reliable binary tracks are not yet available, which could thus underestimate WR populations at later times. Thus, the inferred populations of both WR and O-stars may be altered with different assumptions, which would alter the interpretation.

While consistent with predictions, Figure \ref{fig-sb99:wro} shows S26 may have a somewhat high WR/O ratio, albeit with large uncertainties. While this discrepancy is not statistically significant, there are many physical scenarios that could explain an offset if real. Firstly, stochastic behavior becomes increasingly important for clusters with masses $<10^5$ M$_{\sun}$ \citep{fl10} and S26 is at that limit. A high WR/O ratio could also be produced with a different IMF or a higher upper mass limit than assumed, as in the case of the extreme star formation in NGC 3125  \citep{wof14}. Alternatively, the population ratio could be altered by a multiple burst scenario. Lastly, if S26 has not yet fully emerged from its natal material, some of the stars may be embedded and thus unseen at optical wavelengths. Comparisons of the ionizing flux seen at optical and radio wavelengths show this is unlikely in S26. The optical ionizing flux of S26 is simply estimated through empirical relations from \citet{sv98}. Using H$\beta$ at 4861\AA, the ionizing flux is Q$_0$ $\sim$ 180$\times 10^{49}$ photons s$^{-1}$. The ionizing flux, as measured by the thermal radio emission, can be determined by $$ Q_{\text{Lyc}} \geq 6.3 \times 10^{52} (\frac{T_e}{10^4K}) ^{-0.45} (\frac{\nu}{\text{GHz}})^{0.1} \frac{L_{\nu \text{, thermal}}}{10^{27} \text{erg} \; s^{-1} \;\text{Hz}^{-1}} \;s^{-1} $$ \citep{con92} is Q$_{\text{Lyc}}$ $\sim$ (200$\pm$20) $\times$ 10$^{49}$ s$^{-1}$ at 1.3 cm. The ionizing fluxes agree within the uncertainties, and therefore it is unlikely many O-stars remain embedded.

\subsection{\label{sec-impact} Impact of the Massive Stars on the Cluster Evolution}

\subsubsection{\label{sec-outflow} The Potential Ionized Bipolar Outflow in S26}

\begin{figure}[t]
\includegraphics*[width=12cm,angle=0]{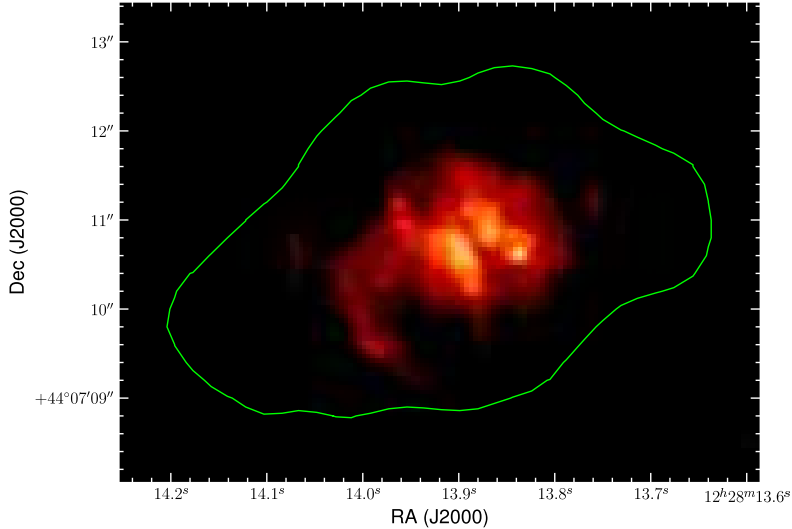}
\caption{\label{fig-bipolar} 
An HST H$\alpha$ image zoomed in on S26, showing the morphology of a bipolar ionized outflow. The green line shows a 3$\sigma$ contour from 3.6 cm continuum, as in Figure \ref{fig-hst}. }

\end{figure}

The morphology of S26 as seen in archival {\em Hubble Space Telescope} imaging with narrow band filters centered on the ionized lines of H$\alpha$, [NIII], and [OIII] suggest that it could be driving an ionized outflow on an intermediate scale, dwarfing individual stellar outflows seen in the Milky Way yet smaller than a galactic outflow. As shown in Figure \ref{fig-bipolar}, the resolved morphology of the central ionized nebular gas in S26 appears bipolar and is evident in all archival HST images taken with ionized gas filters, with a size scale of roughly 1'' $\sim$ 18 pc. The butterfly morphology of S26 most resembles that of bipolar HII regions, which are a type  of UCHII according to the modified classification scheme by \citet{dep05}. As a class, the kinematics of bipolar HII regions suggest an ionized outflow directed by a central source. S26 appears the most morphologically similar to the Galactic bipolar HII region S106, which is surprising given that S106 spans roughly 0.006-0.009 pc across and is ionized by a single O-star  \citep{chu02}, quite different from the massive stellar populations contributing to the nebula in S26 in NGC 4449. Although vastly different spatial scales, both S26 and S106 display a dark lane that bisects the hourglass nebula. In S106, the dark lane is thought to result from a combination of shadowing and protection of an inner disk and high column-density, warm gas on the edge of the molecular cloud \citep{sim12}. However, many factors may contribute to this apparent feature in S26 given its distance and complicated environment.

The intermediate size of possible outflow driven by S26 is quite intriguing if the outflow is confirmed. In the Milky Way, ionized outflows from protostars are commonly on the order of a parsec in size \citep{bac96}. Alternatively, SSCs may contribute galactic winds or outflows that typically span several kpc yet do not appear as localized as S26. The starburst in NGC 1569, including several SSCs, appears to be driving a (uncollimated) massive outflow seen as diffuse X-ray spurs corresponding to well-known H$\alpha$ filaments with a high-velocity expanding component that is over 2 kpc in size \citep{hec95}. Morphologically, S26 is more similar to the bipolar superwind of M82, possibly driven by dense clustering of SSCs \citep[e.g.][]{west07}, or perhaps the superbubble off the nucleus of NGC 3079 that is 1 kpc in diameter and powered either by an AGN or starburst \citep{cec01}.  These outflows are clearly much larger than the ionized region in S26 -- thus If S26 is a verified bipolar outflow, it may provide insight into potential precursors to, or alternatively failed, galactic winds.

\subsubsection{\label{sec-evacuation} The Importance of Winds from Evolved Stars}

For some clusters like S26, it may be possible that the mechanical luminosity due to strong stellar winds of an evolved massive star population is the tipping point in the cluster evacuation process. In dense, high-pressure environments, the combination of typical feedback processes from forming stars (such as radiation pressure) may be insufficient to clear a cluster before a supernova. Yet massive stars, on the main sequence and more evolved, could produce the necessary additional mechanical feedback via wind, in combination with the other forms of feedback, to completely clear a cluster prior to a supernova explosion. However to occur before a supernova event, the removal would have to occur over a very short timescale. It is conceivable the enhanced wind phase of evolved stars could drive this evolutionary transition even more efficiently. Thus, perhaps in certain environments, the question is whether the wind mechanical feedback is required, rather than the dominant feedback mechanism, to clear the cluster before a supernova, which will certainly alter the cluster's ability to survive. 

Massive stars driving cluster evolution via winds have been observed; \citet{gg07} suggest that ELCs are evacuating their surroundings through wind, finding evidence for outflows as the HII gas is not bound. Yet, the extent to which the mechanical luminosity due to stellar winds, particularly those from evolved stars, may contribute has not been fully investigated in the SSC regime, yet it must be substantial. The interaction between the photo-ionized HII region and the stellar wind bubble strongly affect the morphological evolution even in the  case of a single massive star \citep{frey03}. \citet{stt13} show that radiation pressure on a wind-driven shell of a cluster becomes negligible after 3 Myr and highlight the importance of the mechanical feedback. 

The first step in answering whether the winds from evolved stars are an essential component in the feedback scheme driving cluster evolution is to investigate if the massive star habitants of S26 could fully remove its natal material solely via winds. We compare the binding energy of the initial cluster in S26 and to calculations of the energy input by the massive star population, highlighting the additional boost given by the evolved WR stars.
 
We first approximate the binding energy of the cluster as a rough estimate of the energy necessary to remove embedding dust and gas. The region corresponding to the 3.6 cm radio continuum in S26, which includes the nebular emission with the bipolar outflow morphology, is the region being cleared out, called the `core' here. The stellar mass of the core is known \citep[][6.5$\times 10^4  \text{ M}_{\sun}$; ]{rei10}; yet because the dust and gas mass are estimated for S26 over a much larger region, we consider a total core mass produced by a range of SFEs. We approximate the total mass in this central region by including the gas and dust as twice the stellar mass (a roughly 50\% SFE) and, closer to what is observed, using a SFE of 5\% \citep{wil97}. We find the binding energy of S26 is $E_{\text{bind}} = 8.8 \times 10^{49}$ erg (${10 \text{pc}}/{\text{r}}$) to $ 8.8 \times 10^{51}$ erg (${10 \text{pc}}/{\text{r}}$) respectively, where the radius is normalized to a canonical value of 10 pc. Pressure contributions from the outside ISM beyond S26 contribute a negligible energy threshold to overcome when expanding the cluster - however the large surrounding dust and gas mass have not been considered.
 
We compare the binding energy of the cluster core to the effective energy from the cumulative mechanical luminosity output by the winds from massive stars as L$_{\text{wind}}$ $\sim$0.5 $\dot{M} v_{\infty}^2$. We assume a conservative 1\% efficiency in the transfer of the mechanical luminosity into the surrounding material, interpreted over integrated O-star lifetimes  from simulations of \citet{frey06}. As S26 is $\approx$ 3 Myr old and thus the WR phase has likely just begun, we adopt a conservative timeframe for the WR phase of S26 to  be the average WR phase of 0.3 Myr for an individual star. We adopt typical O-star characteristics of $\dot{M}$ $ = 10^{-6} \; M_{\sun} \; \text{yr}^{-1}$ \citep{mas03} and $v_{\infty} = 2000 $ km s$^{-1}$ \citep{kud00} for an O7V star (along with $\eta_o$ to estimate the general O-star population, as in Section \ref{sec-ostars}). We adopt typical  WR star mass-loss rates and terminal velocity values of WN7 and WC4 stars from \citet{cro07}, and treat any metallicity effects as in previous sections for the WRs and as {M}$_{\text{O}}$ $\propto$ Z$^{0.8}$  for the O-stars \citep{mok07}.

To first order, assuming a single starburst, S26 may have been completely cleared by the O-star winds alone -- although the fact that the cluster is still partially embedded indicates this has not yet happened or suggests the SFE in the core is indeed low. Over 3.1 Myr, the estimated cluster population of 237 O-stars would contribute 8.8$\times 10^{49}$ erg through winds. However, 18 of these massive stars have evolved into WR stars, which contribute through much stronger winds. If we account for increased mechanical luminosity for the number of inferred WRs over the average 0.3 Myr long WR phase of an individual star, the total massive star population would output 9.9$\times 10^{49}$ erg over the same 3.1 Myr -- larger than the binding energy for an initial cluster with a radius of 10 pc with a high SFE. Thus, the massive star population has likely contributed enough mechanical luminosity alone to have fully cleared the cluster, especially when WR contributions are considered. S26 likely has not done so because of the surrounding material or because the SFE is low. We hypothesize the pre-supernova mechanical luminosity is crucial in the evacuation process in this case, especially as the removal is not instantaneous and thus more likely to ensure cluster survival.

\subsubsection{\label{evolution} Describing the Evolutionary Phase of S26}

\begin{figure}[t]
\includegraphics*[width=12cm,angle=0]{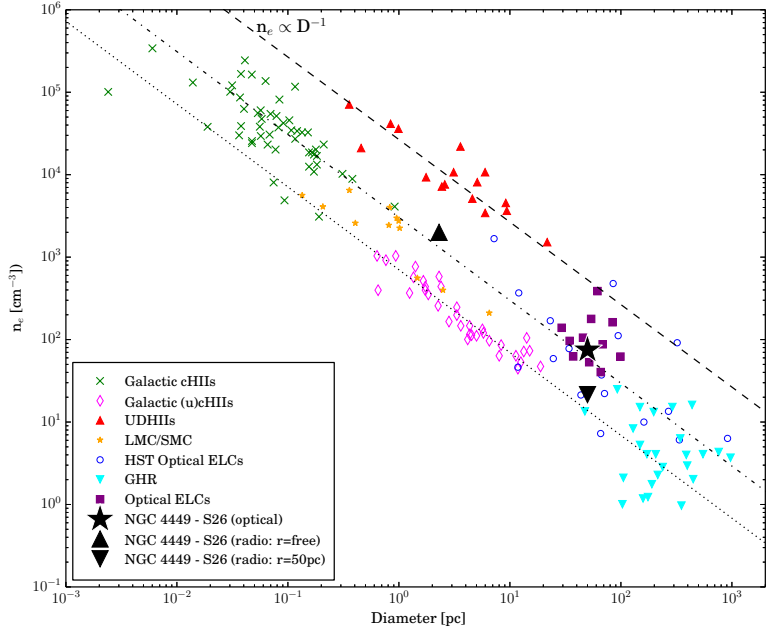}
\caption{\label{fig-sizedensity} 
The optical and radio estimated properties of S26 plotted on the extragalactic HII region size-density relation \citep{gg07,hh09}. As discussed in Section \ref{sec-radiothermal}, the observed radio emission can be modeled with an input radius of 50 pc or with the radius as a free parameter. Color version available online.}

\end{figure}

We have examined S26 to investigate this seemingly short-lived yet critical stage in massive/super star cluster evolution, and we now  identify its place among the major  classes of HII regions. In Figure \ref{fig-sizedensity}, we plot the observed characteristics of S26 on the size-density relation of extragalactic HII regions from \citet{hh09}. The optically derived density, adopting a size of 50 pc as suggested by the 3$\sigma$ contour of the 3.6 cm radio continuum that contains the optical cluster, puts S26 among the ELC regions. The two scenarios consistent with the radio data, described in Section \ref{sec-radiothermal}, show that S26 exhibits characteristics similar to both GHRs or UDHIIs, depending on whether the radius is allowed as a free parameter or set at 50 pc. In the context of the evolutionary classification scheme described by \citet{w14}, S26 lies on the border between Stage 3 (emerging cluster) and Stage 4 (young cluster).

\subsection{\label{sec-lmc} Similarities to 30 Doradus in the LMC}

We compare S26 to 30 Doradus in the LMC located at about 50 kpc \citep{wal12}, whose dense stellar core R136 is the `prototype' core of a SSC \citep{mh98}. In terms of the observed stellar populations, S26 more closely resembles the core R136. The S26 and R136 star forming regions have almost identical stellar masses ($\approx$6$\times$10$^4$ M$_{\sun}$ \citet{h95} for R136 and $\approx$6.5$\times$10$^4$ M$_{\sun}$ \citet{rei10} for S26). S26 hosts about two times the massive star population: R136 contains $\sim$121 massive stars within a 4.7 pc region \citep{h95} and S26 hosts 237 massive stars in a region roughly 50 pc in size. Although R136 may be more concentrated, a tight stellar core of S26 may be unresolved (see Figure \ref{fig-hst}). As discussed in Section \ref{sec-radiothermal}, the radio data in S26 are consistent with thermal emission resulting from both a large, low-density region (constraining a spectra fit to a radius of 50pc) or a small, dense region (2.3 pc; unconstrained radius), and are likely indicative of a low filling factor of high density regions. 

However, much of the core of 30 Dor would be unresolved if at the distance of NGC 4449, and we thus compare the entire systems as well. Within the total 30 Dor region, the VLT-FLAMES Tarantula Survey found 722 massive stars in a 150 pc radius (of which 500 have been spectroscopically confirmed), resulting in a stellar mass of 1.1 $\times10^{5}$ M$_{\sun}$  \citep{dor13} for 30 Dor itself. The scaled stellar mass of S26 is estimated to be 3.0  $\times10^{5}$ M$_{\sun}$ in the large 300 pc aperture, which would likewise host $\sim$1000 massive stars following the same scaling. Most importantly,  30 Dor contains a remarkably similar gas mass of (1.3 $\pm$ 0.5)$\times10^{7}$ M$_{\sun}$ \citep{kim03} over an aperture of 200 pc, which roughly matches the gas mass of 3$\times10^{7}$ M$_{\sun}$ of S26 with the large aperture of 300 pc. 

30 Dor proves an interesting comparison to S26 as it is close enough to resolve individual components and evaluate the impact of the massive star populations on the surrounding environment. Out of the 500 confirmed hot luminous stars in the VLT-FLAMES Tarantula Survey, 31 are WR or Of stars that contribute $\sim$50\% of the wind luminosity and $\sim$40\% of the ionizing luminosity in the cluster \citep{dor13}. This region in 30 Dor roughly corresponds to the size sampled by the small aperture SED of S26 ($\sim$100 pc), and although the optical spectra were obtained for only the inner 50 pc of S26, the estimated evolved population of 18 WRs (out of 237 massive stars) is similar. Thus, the wind and ionizing luminosity can be extrapolated from the resolved 30 Dor case and assumed to be just as significant, if not more so, for S26.

Comparing S26 and 30 Dor can shed light on how feedback mechanisms may become relevant, especially later in evolution. Comparisons of the observed pressure due to stellar radiation, shock-heated hot gas, warm ionized gas (HII gas), and dust processed IR radiation indicate that radiation pressure dominates within the central 75 pc region of R136, and HII gas pressure dominates beyond that \citep{lop11}. Another study that paired observations with photoionization models suggests that the hot X-ray gas instead dominates the mechanics of 30 Dor, and that radiation pressure does not currently play a major role in the structure \citep{pel11}. The importance of radiation pressure in at least the initial expansion is clear, however these and many other studies do not directly include mechanical wind feedback from massive stars.

While 30 Dor has undergone a more complicated star formation history and likely undergone multiple bursts of star formation over the last 20 Myr \citep{dm11}, several parameters of the 30 Dor region parallel those in S26: for one, the gas reservoirs are roughly the same size, stellar feedback is ongoing, and mechanical feedback is largely due to the most massive and evolved stars. The structure of 30 Dor is naturally better defined, with the core hosting 121 massive stars ($\sim$5 Wolf-Rayet), with many more in the vicinity, and is consistent with the picture that the most massive stars curtail further star formation \citep{mh98}. S26 hosts 237 massive stars (18 Wolf-Rayet) and appears to be undergoing or ending the first star formation event as it is $\sim$ 3 Myr old.  Thus, it seems plausible that S26 is a younger version of 30 Dor.

\section{\label{sec-conclusions} Summary}

We have presented a detailed analysis of the partially embedded massive star cluster S26 in NGC 4449. The main results are as follows:

\begin{enumerate}

\item We estimate S26 hosts massive star populations of roughly 18 Wolf-Rayet stars and 219 O-stars from optical spectral lines by assuming standard WR optical line fluxes for a single WR star and a typical ionizing flux for an O-star.

\item The massive star population ratio comparing the subtypes WN/WC in S26 is consistent with predictions and other observed clusters and galaxies. The population ratio WR/O is also consistent with predictions. If unavoidably large uncertainties are ignored, S26 is among the high WR/O ratios observed for other individual clusters of intense star formation.

\item Partially embedded by dust, S26 is one of the dominating sources of the IR emission in NGC 4449. The best fit model to the large aperture SED (300 pc aperture) suggests PAH emission that is stronger than the NGC 4449 galactic value. 

\item Dust model fitting to the infrared SED extracted from 100 pc and 300 pc apertures suggest the exciting radiation increases and PAH emission decreases towards the cluster center. This radial trend suggests the dust grains are being destroyed from within, likely by the massive star feedback. 

\item As estimated by the infrared photometry and the best fit model to the large aperture SED, the total dust mass of the S26 structure is 2.3$\times10^{5}$ M$_{\sun}$; by assuming a dust-to-gas ratio of 130, we estimate a gas mass of 3$\times10^{7}$ M$_{\sun}$. This results in a low star formation efficiency of 1\% over the large 300 pc aperture.

\item We hypothesize that the mechanical luminosity from the evolved stellar winds in S26 may be essential to the emerging process for this cluster, which may ultimately influence future cluster survival. Resolved {\em HST} images of the ionized gas in S26 display an hourglass nebula, which may suggest a possible bipolar ionized outflow. Simple energy calculations suggest that the winds from massive stars may be sufficient in clearing the cluster (particularly if the star formation efficiency is high in the region being cleared). However, as S26 has not been fully evacuated at $\sim$ 3 Myr, the increased feedback contributed by the Wolf-Rayet stars may be necessary in clearing out the cluster.

\item The optical characteristics of S26 match those of Emission Line Clusters on the size-density relation of extragalactic HII regions. The radio properties of S26 cannot discriminate between the UDHII and GHR extremes, which may imply a low filling factor of high density regions within a large, low density HII region. 

\item Similarities between  the S26 cluster and the super star cluster analog 30 Doradus in the LMC may suggest that S26 is akin to a younger version of 30 Dor.

\end{enumerate}

It is evident that the massive star cluster S26 in NGC 4449 is undergoing an important evolutionary stage in which stellar feedback is particularly important. We propose S26 may be an example of a short-lived, yet important phase in massive and super star cluster evolution,  during which the complete evacuation of natal material is aided by the mechanical luminosity of the massive stars, particularly evolved Wolf-Rayet stars.

\acknowledgments
We thank the anonymous referee for useful comments. K.E.J. gratefully acknowledges support provided by the David and Lucile Packard Foundation. 
Observations at the 3.5 m Telescope at Apache Point Observatory are supported by the F. H. Levinson Fund of the Silicon Valley Community Foundation. Support for A.E.R. was provided by NASA through the Einstein Fellowship Program, grant PF1-120086. This research has utilized NASA's Astrophysics Data System Service.


\begin{thebibliography}{}

\bibitem[Agertz et al.(2013)]{ag13} Agertz, O., Kravtsov, A.V., Leitner, S.N., \& Gnedin, N.Y. 2013, \apj, 770, 25
\bibitem[Aniano et al.(2011)]{ani11} Aniano, G., Draine, B.T., Gordon, K.D., \& Sandstrom, K. 2011, \pasp, 123, 1218
\bibitem[Annibali et al.(2008)]{ann08} Annibali, F., Aloisi, A., Mack, J., Tosi, M., van der Marel, R.P., Angeretti, L., Leitherer, C., \& Sirianni, M. 2008, \apj, 135, 1900
\bibitem[Ashman \& Zepf(2001)]{ash01} Ashman, K.M., \& Zepf, S.E. 2001, \aj, 122, 1888
\bibitem[Asplund et al.(2009)]{asp09} Asplund, M., Grevesse, N., Sauval, A.J., \& Scott, P. 2009, \araa, 47, 481
\bibitem[Aversa et al.(2011)]{av11} Aversa, A.G., Johnson, K.E., Brogan, C.L., Goss, W.M., \& Pisano, D.J. 2011, \aj, 141, 125
\bibitem[Baars et al.(1977)]{baar77} Baars, J.W.M., Genzel, R., Pauliny-Toth, I.I.,K., \& Witzel, A. 1977, \aap, 61, 99
\bibitem[Bachiller (1996)]{bac96} Bachiller, R. 1996, \araa, 34, 111
\bibitem[Baumgardt \& Kroupa(2007)]{bau07} Baumgardt, H., \& Kroupa, P. 2007, \mnras, 380, 1589
\bibitem[Balog et al.(2014)]{bal14} Balog, Z. et al. 2014, ExA , 37, 129
\bibitem[Bibby \& Crowther(2010)]{bib10} Bibby, J.L., \& Crowther, P.A. 2010, \mnras, 405, 2737
\bibitem[Bibby \& Crowther(2012)]{bib12} Bibby, J.L., \& Crowther, P.A. 2012, \mnras, 420, 3019
\bibitem[Cappa, Goss, \& van der Hucht(2004)]{cap04} Cappa, C., Goss, W.M., \& van der Hucht, K.A. 2004, \aj, 127, 2885
\bibitem[Cecil et al.(2001)]{cec01} Cecil, G., Bland-Hawthorn, J., Veilleux, S., \& Fillippenko, A.V. 2001, \apj, 555, 338
\bibitem[Chandar et al.(2004)]{chan04} Chandar, R., Leitherer, C., \& Tremonti, C.A. 2004, \apj, 604, 153
\bibitem[Churchwell(2002)]{chu02} Churchill, E. 2002, \araa, 40, 27
\bibitem[Condon (1992)]{con92} Condon, J. \araa, 30, 575
\bibitem[Conti(1991)]{con91} Conti, P.S. 1991, \apj, 377, 115
\bibitem[Conti(1993)]{con93} Conti, P.S. 1993, ASPC, 35, 449
\bibitem[Crowther \& Bibby(2009)]{cro09} Crowther, P.A., \& Bibby, J.L. 2009, \aap, 499, 455
\bibitem[Crowther(2007)]{cro07} Crowther, P.A. 2007, \araa, 24, 177
\bibitem[Crowther \& Hadfield(2006)]{cro06} Crowther, P.A., \& Hadfield, L.J. 2006, \aap, 449, 711
\bibitem[De Marchi et al.(2011)]{dm11} De Marchi, G. et al. 2011, \apj, 739, 27
\bibitem[De Pree et al.(2005)]{dep05} De Pree, C.G., Wliner, D.J., Deblasio, J., Mercer, A.J., \& Davis, L.E. 2005, \apj, 624L, 101
\bibitem[De Robertis et al.(1987)]{der87} De Robertis, M.M., Dufour, R.J., \& Hunt, R.W. 1987, \jrasc, 81, 195
\bibitem[Doran et al.(2013)]{dor13} Doran, E.I. et al. 2013 , \aap, 558, 134
\bibitem[Draine \& Li(2007)]{draine07} Draine, B.T., \& Li, A. 2007, \apj, 657, 810
\bibitem[Engelbracht et al.(2007)]{eng07} Engelbracht, C.W. et al. 2007, \pasp, 119, 994
\bibitem[Fazio et al.(2004)]{faz04} Fazio, G.G., Hora, J.L., Allen, L.E. et al. 2004, \apjs, 154, 10
\bibitem[Fitzpatrick(1985)]{fitz85} Fitzpatrick, E.L. 1985, \apj, 299, 219
\bibitem[Fouesneau \& Lan\c{c}on(2010)]{fl10} Fouesneau, M. \& Lan\c{c}on, A. 2010, \aap, 521, 22
\bibitem[Freyer et al.(2003)]{frey03} Freyer, T., Hensler, G., \& Yorke, H.W. 2003, \apj, 594, 888
\bibitem[Freyer et al.(2006)]{frey06} Freyer, T., Hensler, G., \& Yorke, H.W. 2006, \apj, 638, 262
\bibitem[Fritz et al.(2012)]{fri12} Fritz, H. et al. 2012, \aap, 546,A34
\bibitem[Georgy et al.(2012)]{geo12} Georgy, C., Ekstr\"{o}m, S., Meynet, P., Levesque, E.M., Hirschi, R., Eggenberger, P., \& Maeder, A. 2012, \aap, 542, A29
\bibitem[Gilbert \& Graham(2007)]{gg07} Gilbert, A.M., \& Graham, J.R. 2007, \apj, 668, 168
\bibitem[Gonz\'{a}lez Delgado et al.(2002)] {gd02}  Gonz\'{a}lez Delgado, R.M., Leitherer, C., Stasi\'{n}ska, G., \& Heckman, T.M. 2002, \apj, 580, 824
\bibitem[Gonz\'{a}lez Delgado et al.(1997)] {gd97}  Gonz\'{a}lez Delgado, R.M., Leitherer, C., Heckman, T.M., \& Cervi\~{n}o, M., 1997, \apj, 483, 705
\bibitem[Griffin et al.(2010)]{grif10} Griffin, M.J., Abergel, A., Abreu, A. et al. 2010, \aap, 518, L3
\bibitem[Guseva, Izotov, \& Thuan(2000)]{gus00} Guseva, N.G., Izotov, Y.I, \& Thuan, T.X. 2000, \apj, 531, 776
\bibitem[Hadfield \& Crowther(2006)]{had06} Hadfield, L.J., \& Crowther, P.A. 2006, \mnras, 368, 1822
\bibitem[Hadfield \& Crowther(2007)]{hc07} Hadfield, L.J., \& Crowther, P.A. 2007, \mnras, 381, 418
\bibitem[Hadfield \& Crowther(2008)]{had08} Hadfield, L.J., \& Crowther, P.A. 2008, IAUS, 250, 327
\bibitem[Heckman et al.(1995)]{hec95} Heckman, T.M., Dahlem, M., Lehnert, M.D., Fabbiano, G., Gilmore, D., \& Waller, W.H.1995, \apj, 448, 98
\bibitem[Henry et al.(2007)]{hen07} Henry, A.L., Turner, J.L., Beck, S.C., Crosthwaite, L.P., \& Meier, D.S. 2007, \aj, 133, 757
\bibitem[Hunt \& Hirashita(2009)]{hh09} Hunt, L.K., \& Hirashita, H. 2009, \aap, 507, 1327
\bibitem[Hunter et al.(1995)]{h95} Hunter, D.A., Shaya, E.J., Holtzman, J.A., Light, R.M., O'Neil, E.J., Jr., \& Lynds, R. 1995, \apj, 448, 179
\bibitem[Izotov et al.(1994)]{izo94} Izotov, Y.I., Thuan, T.X., \& Lipovetsky, V.A. 1994, \apj, 435, 647
\bibitem[Izotov et al.(1997)]{izo97} Izotov, Y.I., Thuan, T.X., \& Lipovetsky, V.A. 1997, \apjs, 108, 1
\bibitem[Izotov et al.(2006)]{izo06} Izotov, Y.I., Stasi\'{n}ska, G., Meynet, G., Guseva, N.G., \& Thuan, T.X. 2006, \aap, 448, 955
\bibitem[Johnson(2002)]{john02} Johnson 2002, Science, 297, 776
\bibitem[Johnson \& Kobulnicky(2003)]{jk03} Johnson, K.E. \& Kobulnicky, H.A. 2003, \apj, 597, 923
\bibitem[Johnson et al.(2004)]{john04b} Johnson, K.E., Indebetouw, R., Watson, C., \& Kobulnicky, H.A. 2004, \aj, 128, 610 
\bibitem[Johnson et al.(2009)]{john09} Johnson, K.E., Hunt, L.K., \& Reines, A.E. 2009, \aj, 137, 3788
\bibitem[Karczewski et al.(2013)]{kar13} Karczewski, O.\L., Barlow, M.J., Page, M.J. et al. 2013, \mnras, 431, 2493 
\bibitem[Kepley et al.(2014)]{kep14} Kepley, A.A., Reines, A.E., Johnson, K.E., \& Walker, L.M. 2014, \aj, 147, 43
\bibitem[Kim et al.(2003)]{kim03} Kim, S., Staveley-Smith, L., Dopita, M.A., Sault, R.J., Freeman, K.C., Lee, Y., \& Chu, Y.-H. 2003, \apjs, 148, 473
\bibitem[Kobulnicky \& Johnson(1999)]{kj99} Kobulnicky \& Johnson 1999, ApJ, 527, 154
\bibitem[Kroupa et al.(2001)]{kro01} Kroupa, P., Aarseth, S., \& Hurley, J. 2001, \mnras, 321, 699
\bibitem[Kudritzki \& Puls(2000)]{kud00} Kudritzki, R.-P., \& Puls, J. 2000, \araa, 38, 613
\bibitem[Leitherer(1999)]{lei99} Leitherer, C. 1999, IAUS, 193, 526
\bibitem[Leitherer et al.(1999)]{sb99} Leitherer, C. et al. 1999, \apjs, 123, 3
\bibitem[Lequeux et al.(1979)]{leq79} Lequeux, J., Peimbert, M., Rayo, J.F., Serrano, A., \& Torres-Peimbert, S. 1979, \aap, 80, 155
\bibitem[Lopez et al.(2011)]{lop11} Lopez, L.A., Krumholz, M.R., Bolatto, A.D., Prochaska, J.X., \& Ramirez-Ruiz, E. 2011, \apj, 731,91 
\bibitem[Lopez et al.(2013)]{lop13} Lopez, L.A., Krumholz, M.R., Bolatto, A.D., Prochaska, J.X., Ramirez-Ruiz, E., \& Castro, D. 2014, \apj, 795, 121
\bibitem[Madden et al.(2013)]{mad13} Madden, S.C., R\'{e}my-Ruyer, A., Galametz, M. et al. 2013, \pasp, 125, 600
\bibitem[Maeder \& Conti(1994)]{mae94} Maeder, A., \& Conti, P.S. 1994, \araa, 32, 227
\bibitem[Martins et al. (2005)]{mart05} Martins, F., Schaerer, D., \& Hiller, D.J. 2005, \aap, 436, 1049
\bibitem[Massey(1996)]{mas96} Massey, P. 1996, Liege International Astrophysical Colloquia, ed. Vreux, J.~M., Detal, A., Fraipont-Caro, D., Gosset, E., \& Rauw, G. 33, 361
\bibitem[Massey(2003)]{mas03} Massey, P. 2003, \araa, 41, 15
\bibitem[Massey \& Holmes(2002)]{mas02} Massey, P., \& Holmes, S. 2002, \apj, 580, L35
\bibitem[Massey \& Hunter(1998)]{mh98} Massey, P., \& Hunter, D.A. 1998, \apj, 493, 180
\bibitem[Massey \& Johnson(1998)]{mj98} Massey, P., \& Johnson, O. 1998, \apj, 505, 793
\bibitem[Mathis et al.(1983)]{mat83} Mathis, J.S., Mezger, P.G., \& Panagia, N. 1983, \aap, 128, 212
\bibitem[Meynet \& Maeder(2005)]{mm05} Meynet, G., \& Maeder, A. 2005, \aap,429, 581
\bibitem[Misselt, Clayton, \& Gordon(1999)]{mis99} Misselt, K.A., Clayton, G.C., \& Gordon, K.D. 1999, \apj, 515, 128
\bibitem[Mokiem et al.(2007)]{mok07} Mokiem, M.R., et al. 2007, \aap, 473, 603
\bibitem[Moll et al.(2007)]{mol07} Moll, S.L., Mengel, S., De Grijs, R., Smith, L.J., \& Crowther, P.A. 2007, \mnras,382,1877 
\bibitem[Meurer et al.(1992)]{meu92} Meurer, G.R., Freeman, K.C., Dopita, M.A., \& Cacciari, C. 1992, \aj, 103, 60
\bibitem[Neugent et al.(2012)]{neu12} Neugent, K.F., Massey, P., \& Georgy, C. 2012, \apj, 759, 11
\bibitem[Neugent \& Massey(2011)]{neu11} Neugent, K.F., \& Massey, P. 2011, \apj, 733, 123
\bibitem[Pellegrini et al.(2011)]{pel11} Pellegrini, E.W., Baldwin, J.A., \& Ferland, G.J. 2011, \apj, 738, 34
\bibitem[Pfalzner \& Kaczmarek(2013)]{pfa13} Pfalzner, S. \& Kaczmarek, T. 2013, \aap, 559A, 38
\bibitem[Pilbratt et al.(2010)]{pil10} Pilbratt, G.L., Riedinger, J.R., Passvogel, T. et al. 2010, \aap, 518, L1
\bibitem[Poglitsch et al.(2010)]{pog10} Poglitsch, A., Waelkens, C., Geis, N. et al. 2010, \aap, 518, L2
\bibitem[Reach et al.(2005)]{rea05} Reach, W.T. et al. 2005, \pasp, 117, 978
\bibitem[Reines, Johnson, \& Goss(2008)]{rei08} Reines, A.E., Johnson, K.E., \& Goss, W.M. 2008, \aj, 135, 2222
\bibitem[Reines et al.(2010)]{rei10} Reines, A.E., Nidever, D.L., Whelan, D.G., \& Johnson, K.E. 2010, \apj, 708, 26
\bibitem[Rela\~{n}o et al.(2013)]{rel13} Rela\~{n}o, M. et al. 2013, \aap, 522, A140
\bibitem[R\'{e}my-Ruyer et al.(2013)]{rr13} R\'{e}my-Ruyer, A. et al. 2013, \aap, 557, 95 
\bibitem[Rieke et al.(2004)]{riek04} Rieke, G., and E.T. Young, C.W. Engelbracht, et al. 2004, \apjs, 154, 25
\bibitem[Rogers \& Pittard(2013)]{rp13} Rogers, H., \& Pittard, J.M. 2013, \mnras, 431, 1337
\bibitem[Russell \& Dopita(1990)]{rus90} Russell, S.C., \& Dopita, M.A. 1990, \apjs, 74, 93
\bibitem[Ry\'{s} et al.(2011)]{rys11} Ry\'{s}, A., Grocholski, A.J., van der Marel, R.P., Aloisi, A., \& Annibali, F. 2011, \aap, 530, A23
\bibitem[Schaerer, Contini, \& Pindao(1999)]{sch99} Schaerer, D., Contini, T., \& Pindao, M. 1999, \aap SS, 136, 35
\bibitem[Schaerer \& Vacca(1998)]{sv98} Schaerer, D., \& Vacca, W.D. 1998, \apj, 497, 618
\bibitem[Schlegel, Finkbeiner, \& Davis(1998)]{sch98} Schlegel, D.J., Finkbeiner, D.P., \& Davie, M. 1998, \apj, 500, 525
\bibitem[Sidoli et al.(2004)]{sid04} Sidoli, F., Smith, L.J., Crowther, P.A., Vacca, W.D., \& Schmutz, W. 2004, ASP Conf. Ser., ed. Lamers, H.J.G.L.M.,Smith, L.J., \& Nota, A. 322
\bibitem[Sidoli et al.(2006)]{sid06} Sidoli, F., Smith, L.J., \& Crowther, P.A. 2006, \mnras, 370, 799
\bibitem[Silich \& Tenorio-Tagle(2013)]{stt13} Silich, S., \& Tenorio-Tagle, G. 2013, \apj, 765, 43
\bibitem[Simon et al.(2012)]{sim12} Simon, R.,et al. 2012, \aap, 542, L12
\bibitem[Smith, Norris, \& Crowther(2002)]{smi02} Smith, L.F., Norris, R.P.F., \& Crowther, P.A. 2002, \mnras, 337, 1309
\bibitem[Stasi\'{n}ska(1990)]{sta90} Stasi\'{n}ska, G. 1990, \aaps, 83, 501
\bibitem[Telles et al.(2014)]{tel14} Telles, E., Thuan, T.X., Izotov, Y.I., \& Carrasco, E.R. 2014, \aap, 561, A64
\bibitem[Tsai et al.(2006)]{tsai06} Tsai, C.-W., Turner, J.L., Beck,S.C., Crosthwaite, L.P., Ho, P.T.P., \& Meier, D.S. 2006, \aj, 132, 2383
\bibitem[Tsai et al.(2009)]{tsai09} Tsai, C.-W., Turner, J.L., Beck, S.C., Meier, D.S., \& Ho, P.T.P. 2009, \aj, 137, 4655
\bibitem[Turner et al.(2003)]{turn03} Turner, J.L., Beck, S.C., Crosthwaite, L.P., Larkin, J.E., McLean, I.S., \& Meier, D.S., Nature, 423, 621
\bibitem[Turner et al.(2000)]{turn00}Turner, J.L., Beck, S.C., \& Ho 2000, \apj, 532, 109
\bibitem[Walker(2012)]{wal12} Walker, A.R. 2012, \apss, 341, 43
\bibitem[Werner et al.(2004)]{wer04} Werner, M., Roellig, T., Low, F. et al. 2004, \apjs,  154, 1
\bibitem[Westmoquette et al.(2007)]{west07} Westmoquette, M.S., Smith, L.J., Gallagher III, J.S., O'Connell, R.W., Rosario, D.J., \& De Grijs, R. 2007, \apj, 671, 358
\bibitem[Whitmore et al.(2014)]{w14} Whitmore, B.C. et al. 2014, \apj, 795, 156
\bibitem[Whitmore \& Schweizer(1995)]{ws95} Whitmore, B.C. \& Schweizer, F. 1995, \aj, 109, 960
\bibitem[Williams \& McKee(1997)]{wil97} Williams, J.P., \& McKee, C.F. 1997, \apj, 476, 166
\bibitem[Wilson et al.(2008)]{wil08} Wilson, C.D., et al. 2008, \apjs, 178, 189
\bibitem[Wofford et al.(2014)]{wof14} Wofford, A., Leitherer, C., Chandar, R., \& Bouret, J.-C. 2014, \apj, 781, 122
\end{thebibliography}
\end{document}